\documentclass[submission,Phys]{SciPost}

\usepackage[utf8]{inputenc} % input encodings (allow UTF-8 input)
\usepackage[T1]{fontenc} 	% selecting font encodings (use 8-bit T1 fonts)
\usepackage[english]{babel} % multilingual support (English language/hyphenation)

\usepackage[bitstream-charter]{mathdesign}
\usepackage{geometry} 		% flexible and complete interface to document dimensions
\usepackage{amsmath} 		% math package (American Mathematical Society)
\usepackage{mathtools} 		% math package (fixes various deficiencies of amsmath)
\usepackage{float} 			% floating objects such as figures and tables
\usepackage{graphicx} 		% enhanced support for graphics
\usepackage{tabularx} 		% tabulars with adjustable-width columns
\usepackage{booktabs} 		% professional-quality tables
\usepackage{color, xcolor} 	% foreground and background colour management
\usepackage{pdfpages} 		% inclusion of external multi-page PDF documents
\usepackage{extarrows} 		% extra arrows beyond those provided in amsmath
\usepackage{multirow} 		% create tabular cells spanning multiple rows
\usepackage{multicol} 		% intermix single and multiple columns
\usepackage{caption} 		% customising captions in floating environments
\usepackage{subcaption} 	% support for sub-captions (captions for subfigures)
\usepackage{enumitem} 		% control layout of itemize, enumerate, description
\usepackage{xspace} 		% define commands that appear not to eat spaces
\usepackage{stackrel} 		% enhancement to the \stackrel command
\usepackage{tikz} 			% create PostScript and PDF graphics
\usetikzlibrary{calc}
\usepackage{braket} 		% Dirac bra-ket notation
\usepackage{bm} 			% access bold symbols in maths mode
\usepackage{tensor} 		% typeset tensors (tensor-style super- and subscripts)
\usepackage{slashed} 		% slash through characters (Feynman slash notation)
\usepackage{siunitx} % SI units package (typesetting values with units)
\usepackage{lastpage} 		% reference last page
\usepackage{cite} 			% improved citation handling
\usepackage[normalem]{ulem} % package for underlining
\usepackage{fontawesome} 	% access to web-related icons
\usepackage{tocloft} 		% control over the typography of the TOC, LOF, and LOT
\usepackage{titlesec} 		% interface to sectioning commands (various title styles)
\usepackage{doi} 			% create correct hyperlinks for DOI numbers
\usepackage[most]{tcolorbox} 					% coloured and framed text boxes
\usepackage[capitalize]{cleveref} 	% intelligent cross-referencing
\usepackage[nottoc, notlot, notlof]{tocbibind} 	% add references/index/contents to TOC
\usepackage[ruled, vlined]{algorithm2e} 		% floating algorithm environment
\usepackage{makecell}
\usepackage{pifont}
\usepackage{stackrel}
\usepackage{feynmp-auto}

\hypersetup{
	pdftitle={Jet Diffusion versus JetGPT — Modern Networks for the LHC},
	pdfauthor={Anja Butter, Nathan Huetsch, Sofia Palacios Schweitzer, Tilman Plehn, Peter Sorrenson and Jonas Spinner},
    linktoc=section,
	breaklinks=true,			% splits links across lines
	colorlinks=true, 			% false: boxed links, true: colored links
	linkcolor={red!50!black}, 	% color of internal links (sections, pages, etc.)
	citecolor={blue!50!black}, 	% color of citation links (links to bibliography)
	urlcolor={blue!80!black} 	% color of URL links (external links)
} 
\DeclareSymbolFont{usualmathcal}{OMS}{cmsy}{m}{n}
\DeclareSymbolFontAlphabet{\mathcal}{usualmathcal}

\SetArgSty{textnormal}
\SetKwComment{Comment}{{\small\#}~}{}
\SetCommentSty{mycommfont}

\setitemize{itemsep=2pt, parsep=0pt} 				% adjust itemize environment
\setenumerate{itemsep=2pt, parsep=0pt} 				% adjust enumerate environment
\crefname{section}{Sec.}{Secs.}
\crefname{equation}{Eq.}{Eqs.}
\crefname{figure}{Fig.}{Figs.}
\crefname{table}{Tab.}{Tabs.}
\Crefname{section}{Section}{Sections}
\Crefname{equation}{Equation}{Equations}
\Crefname{figure}{Figure}{Figures}
\Crefname{table}{Table}{Tables}

\usetikzlibrary{arrows, shapes.geometric, arrows.meta, shapes, decorations.pathreplacing, fit, patterns, patterns.meta}

\definecolor{Rcolor}{HTML}{E99595}
\definecolor{Gcolor}{HTML}{C5E0B4}
\definecolor{Bcolor}{HTML}{9DC3E6}
\definecolor{Ycolor}{HTML}{FFE699}

\tikzstyle{expr} = [circle, minimum width=1.8cm, minimum height=1.8cm, text centered, align=center, inner sep=0, draw,font=\LARGE]
\tikzstyle{txt_huge} = [align=center, font=\Huge, scale=2]
\tikzstyle{txt} = [align=center, font=\LARGE]
\tikzstyle{cinn} = [double arrow, double arrow head extend=0cm, double arrow tip angle=130, shape border rotate=90, inner sep=0, align=center, minimum width=2.1cm, minimum height=2.3cm, fill=Bcolor, draw,font=\LARGE]
\tikzstyle{cinn_black} = [cinn, minimum height=2.5cm, fill=black]
\tikzstyle{arrow} = [thick,-{Latex[scale=1.0]}, line width=0.2mm, color=black]
\tikzstyle{line} = [thick, line width=0.2mm, color=black]
\tikzstyle{loss} = [rectangle, align=center,  minimum width=1.8cm, minimum height=1.5cm,fill=Rcolor,font=\LARGE, rounded corners]
\tikzstyle{xt} = [rectangle, align=center,  minimum width=4cm, minimum height=1.5cm,fill=Gcolor,font=\Large, rounded corners]
\tikzstyle{xts} = [rectangle, align=center,  minimum width=1cm, minimum height=1.5cm,fill=Gcolor,font=\Large, rounded corners] 
\newcommand{\ie}{i.e.\@\xspace} 		% id est (i.e.)
\newcommand{\XLangle}{\Bigl\langle}
\newcommand{\XRangle}{\Bigr\rangle}

\newcommand{\loss}{\mathcal{L}} 	% loss value

\newcommand{\kl}{\operatorname{KL}}

\newcommand{\pmd}{p_\text{model}}

\newcommand{\arXiv}[2][]{%
	\ifthenelse{\equal{#1}{}}%
	{\href{http://arxiv.org/abs/#2}{arXiv:#2}}%
	{\href{http://arxiv.org/abs/#2}{arXiv:#2~[#1]}}}

\def\slashchar#1{\setbox0=\hbox{$#1$}           % set a box for #1
   \dimen0=\wd0                                 % and get its size
   \setbox1=\hbox{/} \dimen1=\wd1               % get size of /
   \ifdim\dimen0>\dimen1                        % #1 is bigger
      \rlap{\hbox to \dimen0{\hfil/\hfil}}      % so center / in box
      #1                                        % and print #1
   \else                                        % / is bigger
      \rlap{\hbox to \dimen1{\hfil$#1$\hfil}}   % so center #1
      /                                         % and print /
   \fi}

\newcommand{\tikznode}[2]{%
\ifmmode%
\tikz[remember picture,baseline=(#1.base),inner sep=0pt] \node (#1) {$#2$};%
\else
\tikz[remember picture,baseline=(#1.base),inner sep=0pt] \node (#1) {#2};%
\fi}

\def\mathswitchr#1{\relax\ifmmode{\text{#1}}\else$\text{#1}$\xspace\fi}
\def\mathswitch#1{\relax\ifmmode#1\else$#1$\xspace\fi}

\graphicspath{{./figs/}}
\begin{document}

\vspace*{-2.5em}
\hfill{}
\vspace*{0.5em}

\begin{center}{\Large 
\textbf{Kicking it Off(-shell) with Direct Diffusion} \\
}\end{center}

\begin{center}
Anja Butter\textsuperscript{1,3},
Tom\'a\v{s} Je\v{z}o\textsuperscript{2},
Michael Klasen\textsuperscript{2},\\
Mathias Kuschick\textsuperscript{2}, 
Sofia Palacios Schweitzer\textsuperscript{1}, and
Tilman Plehn\textsuperscript{1}
\end{center}

\begin{center}
{\bf 1} Institut f\"ur Theoretische Physik, Universit\"at Heidelberg, Germany
\\
{\bf 2}
Institut f\"ur Theoretische Physik, Universität M\"unster, Germany
\\
{\bf 3} LPNHE, Sorbonne Universit\'e, Universit\'e Paris Cit\'e, CNRS/IN2P3, Paris, France
\end{center}

%\begin{center}
%\today
%\end{center}

% For convenience during refereeing: line numbers
%\linenumbers
\vspace{-1cm}
\section*{Abstract}
         {\bf Off-shell effects in large LHC backgrounds are
           crucial for precision predictions and, at the same time,
           challenging to simulate. 
           We present a novel method to transform high-dimensional distributions based on a diffusion neural network and use it to generate a process with off-shell kinematics from the much simpler on-shell one.
           Applied to a toy example of top pair production at LO we
           show how our method
           generates off-shell configurations fast and 
           precisely, while reproducing even challenging on-shell features.}

\vspace{10pt}
\noindent\rule{\textwidth}{1pt}
\tableofcontents\thispagestyle{fancy}
\noindent\rule{\textwidth}{1pt}
%\vspace{10pt}

\clearpage
%%%%%%%%%%%%%%%%%%%%%%%%%%%%%%%%%%%%%%%%%%%%%%%%%%%%%%%%%%%%%%%%%%%%%%%%
\section{Introduction}
\label{sec:intro}

Fast and precise theoretical predictions from first principles are
required for essentially every experimental LHC analysis. This is
especially true for modern inference methods, where the complete phase
space coverage ensures an optimal
measurement~\cite{Butter:2022rso,Cranmer:2019eaq}.  Based on
perturbative quantum field theory, precision simulations of the hard
scattering process face two challenges, the loop order and the
multiplicity of the partonic final state~\cite{Campbell:2022qmc}.  The
latter increases rapidly for example, when we describe the production of decaying
heavy particles.  Naively, one could expect that describing the decay
kinematics close to the mass shell of the heavy particles, for example
using a Breit-Wigner propagator, is sufficient. However, given the
precision targets of the upcoming LHC runs, on-shell approximations
are no longer
justified~\cite{Heinrich:2017bqp,FerrarioRavasio:2018whr}. In view of
the HL-LHC, simulations need to describe heavy particle production and
decay including, both, quantum corrections and off-shell kinematics.

In addition to the theoretical and computational effort behind
precise calculations, the question is how they can be made available through multi-purpose event
generators. Here we can resort to modern machine learning
(ML)~\cite{Butter:2022rso,Plehn:2022ftl}. Following the modular
structure of event generators, the most obvious ML-applications are
neural network surrogates for expensive scattering
amplitudes~\cite{Bishara:2019iwh,Badger:2020uow,Aylett-Bullock:2021hmo,Maitre:2021uaa,Winterhalder:2021ngy,Badger:2022hwf,Maitre:2023dqz}.
Aside from the speed, these surrogates have the advantage that
they can always be evaluated in parallel on
GPUs~\cite{Hagiwara:2013oka,Bothmann:2021nch,Carrazza:2021gpx,Valassi:2021ljk}.
Given these amplitudes, we can then improve the phase-space
integration and
sampling~\cite{Danziger:2021eeg,Chen:2020nfb,Gao:2020vdv,Bothmann:2020ywa,Gao:2020zvv,Heimel:2022wyj,Heimel:2023ngj,Janssen:2023ahv,Bothmann:2023siu}.
One advantage is that simulations are defined over interpretable physics
phase spaces, for example scattering
events~\cite{dutch,gan_datasets,DijetGAN2,Butter:2019cae,Alanazi:2020klf,Butter:2021csz,Butter:2023fov},
parton
showers~\cite{locationGAN,Andreassen:2018apy,Bothmann:2018trh,Dohi:2020eda,Buhmann:2023pmh,Leigh:2023toe,Mikuni:2023dvk,Buhmann:2023zgc},
and detector
simulations~\cite{Paganini:2017hrr,deOliveira:2017rwa,Paganini:2017dwg,Erdmann:2018kuh,Erdmann:2018jxd,Belayneh:2019vyx,Buhmann:2020pmy,Buhmann:2021lxj,Krause:2021ilc,
  ATLAS:2021pzo,Krause:2021wez,Buhmann:2021caf,Chen:2021gdz,
  Mikuni:2022xry,ATLAS:2022jhk,Krause:2022jna,Cresswell:2022tof,Diefenbacher:2023vsw,
  Hashemi:2023ruu,Xu:2023xdc,Diefenbacher:2023prl,Buhmann:2023bwk,Buckley:2023rez,Diefenbacher:2023flw
}. All of these speed improvements maintain the first-principles
nature of LHC event generators, which means they
take theory predictions and evaluate them using faster ML-methods. 

The goal of this study is to develop a precise and fast generative network 
that maps two LHC phase spaces onto each other~\cite{Golling:2023mqx,Diefenbacher:2023wec,Diefenbacher:2023flw}.
If one phase space is a narrow sub-manifold of the other one, like in the case of on-shell and off-shell phase spaces of production of heavy unstable particles, one cannot use regression or classifier reweighting but must resort to generative networks.
For
LHC applications, we already know that precision-generative
networks~\cite{Butter:2021csz,
  Winterhalder:2021ave,Nachman:2023clf,Leigh:2023zle,Das:2023ktd} are
easy to ship and powerful in amplifying their training
data~\cite{Butter:2020qhk,Bieringer:2022cbs}.  Tasks we can solve with
their help include event subtraction~\cite{Butter:2019eyo}, event
unweighting~\cite{Verheyen:2020bjw,Backes:2020vka}, or
super-resolution enhancement~\cite{DiBello:2020bas,Baldi:2020hjm}.
Their conditional versions enable new analysis methods, like
probabilistic
unfolding~\cite{Datta:2018mwd,Bellagente:2019uyp,Andreassen:2019cjw,Bellagente:2020piv,Backes:2022vmn,Leigh:2022lpn,Raine:2023fko,Shmakov:2023kjj,Ackerschott:2023nax,Diefenbacher:2023wec},
inference~\cite{Bieringer:2020tnw,Butter:2022vkj,Heimel:2023mvw}, or
anomaly
detection~\cite{Nachman:2020lpy,Hallin:2021wme,Raine:2022hht,Hallin:2022eoq,Golling:2022nkl,Sengupta:2023xqy}. Currently, the
best-performing network architectures are normalizing flows for
invertible tasks and diffusion networks for sampling problems, with
help from transformers for combinatorics.

In this paper we present such a mapping and use it to efficiently generate events over an
off-shell phase space from given on-shell events. The idea behind this 
sampling from on-shell events is that the
generative network does not have to reproduce the on-shell features
and can focus on the additional
and relatively smooth off-shell extension. We start by describing our
training dataset and the problem of off-shell event generation in
Sec.~\ref{sec:data}. Our generative network setup, based on a
Bayesian- Conditional Flow Matching (CFM) architecture~\cite{Butter:2023fov}, is
presented in Sec.~\ref{sec:diff}. To control the generative network
performance~\cite{Das:2023ktd} and to improve the precision of the
kinematic distributions~\cite{Butter:2021csz}, we apply a classifier
reweighting in Sec.~\ref{sec:reweight}. We present an Outlook in
Sec.~\ref{sec:outlook}.

%%%%%%%%%%%%%%%%%%%%%%%%%%%%%%%%%%%%%%%%%%%%%%%%%%%%%%%%%%%%%%%%%%%%%%%%
\section{Off-shell vs. on-shell events}
\label{sec:data}

Our benchmark process is the complete off-shell top pair production
followed by leptonic decays,
\begin{align}
 p p \to b e^+ \nu_e \; \bar{b} \mu^- \nu_\mu \; .
\label{eq:proc}
\end{align}
Top pair production with its rich resonance structure is known to
challenge phase space sampling, just as generative networks for
transition amplitudes~\cite{Butter:2019cae}.

In the factorized approach, in which the production and decay
processes decouple, this process is known all the way up to
NNLO-QCD~\cite{Czakon:2013goa,
  Czakon:2015owf,Catani:2019iny,Catani:2019hip},
NLO-EW~\cite{Bernreuther:2008md, Kuhn:2006vh, Hollik:2011ps,
  Gutschow:2018tuk, Frederix:2021zsh} and NNLO-QCD combined with
NLO-EW~\cite{Czakon:2017wor} in the production process;
NNLO-QCD~\cite{Gao:2012ja, Brucherseifer:2013iv} in the decay; and
NNLO-QCD in both production and
decay~\cite{Gao:2017goi,Behring:2019iiv,Czakon:2020qbd}. Full
off-shell calculations in the dilepton channel are so far only
available at NLO-QCD~\cite{Bevilacqua:2010qb, Denner:2010jp,
  Denner:2012yc, Heinrich:2013qaa, Frederix:2013gra,
  Cascioli:2013wga}, but a calculation of full off-shell top pair
production in association with an extra jet at NLO is also available
in Ref.~\cite{Bevilacqua:2015qha}.  In this pilot study we restrict
ourselves to leading order in QCD, $\mathcal{O}(\alpha_S^2 \alpha^4)$,
and reserve the application of our method to higher-order predictions
to a future publication.

We generate event samples for 13~TeV proton-proton collisions with
NNPDF31\_nlo\_as\_0118 parton distributions~\cite{NNPDF:2017mvq}.  The
neutrinos, charged leptons and quarks of the first two generations are
treated as massless, and the CKM matrix is assumed to be trivial.  All
input parameters are given in Tab.~\ref{tab:gen_params}, and the
electromagnetic coupling $\alpha$ and the weak mixing angle are
derived from the weak gauge-boson masses and the Fermi constant.

Our study and our results can be extended to higher-order
predictions or other processes by combining different jet
multiplicities in the final state~\cite{Heimel:2023mvw}.  Our two
benchmark datasets are generated with
\textsc{Hvq}~\cite{Frixione:2007nw} and
\textsc{Bb4l}~\cite{Jezo:2016ujg, Jezo:2023rht}, respectively.  The
data generated with \textsc{Hvq} only includes approximate off-shell
effects using a finite top width and including spin correlations~\cite{Frixione:2007zp} and is
referred to as on-shell data. The generator is based on the
\textsc{Powheg} method~\cite{Nason:2004rx, Frixione:2007vw} and is
part of the \textsc{Powheg\,Box\,V2}~\cite{Alioli:2010xd} package.

The data generated with \textsc{Bb4l} takes into account full
off-shell effects also including singly-resonant and non-resonant contributions and the corresponding interferences. The generator employs the \textsc{PowhegRes}
method~\cite{Jezo:2015aia}, tailored for simulations with unstable
particles.  In this case, $W$ bosons and $b$ quarks do not always stem
from a top decay.\medskip

%----------------------------------------------------------
\begin{table}[b!]
\centering \begin{small} \begin{tabular}{l|cc|l|c}
\toprule
    $m_t$  & $172.5 \; \si{GeV}$ && $\Gamma_t$ & $1.453 \; \si{GeV}$  \\
    $m_b$  & $4.75 \; \si{GeV}$ &&&\\
    $m_Z$  & $91.188 \; \si{GeV}$&& $\Gamma_Z$& $2.441 \; \si{GeV}$\\
    $m_W$  & $80.419 \; \si{GeV}$ && $\Gamma_W$ & $2.048 \; \si{GeV}$ \\
    $m_H$  &$125.0 \; \si{GeV}$&&$\Gamma_H$&$0.0403 \;\si{GeV}$ \\
    \midrule
    \multicolumn{3}{c|}{$\mathcal{B}(W \to e \nu/\mu \nu)$}  & \multicolumn{2}{c}{1/9} \\ \multicolumn{3}{c|}{$G_F$} & \multicolumn{2}{c}{$1.16585 \times 10^{-5} \si{GeV}^{-2}$} \\ 
    \bottomrule
\end{tabular}
\end{small}
\caption{Parameters used for the generation of the training datasets.}
\label{tab:gen_params}
\end{table}
%----------------------------------------------------------

\begin{figure}[t]
    \includegraphics[width=0.495\textwidth,page=60]{data_plots/Observables_off_on}
    \includegraphics[width=0.495\textwidth,page= 3]{data_plots/Observables_off_on}\\
    \includegraphics[width=0.495\textwidth,page=58]{data_plots/Observables_off_on}
    \includegraphics[width=0.495\textwidth,page= 1]{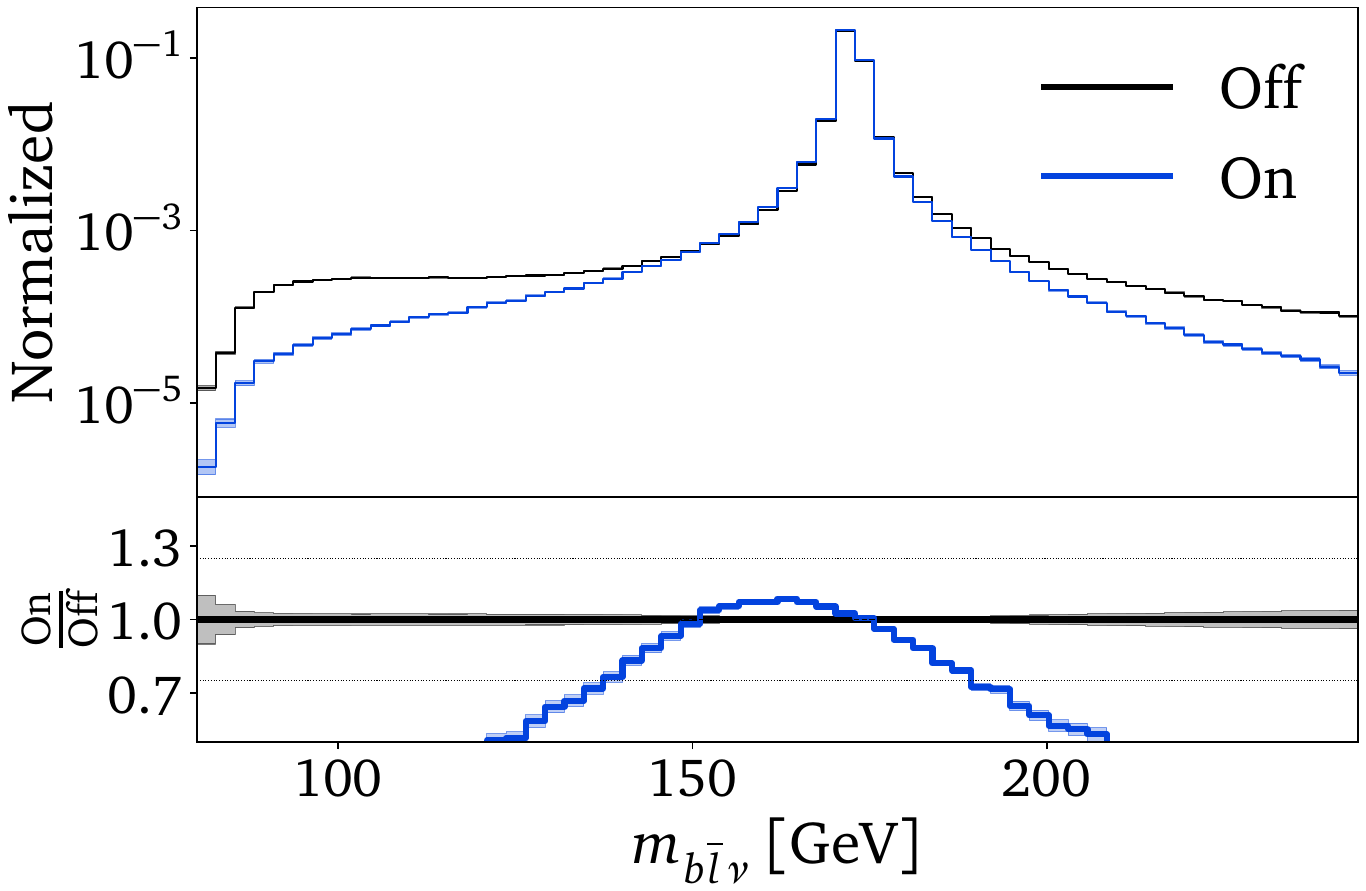}
    \caption{Example distributions for on-shell and off-shell
      $t\bar{t}$ event samples, illustrating the different phase space
      coverage. The right panels show the same distribution as the
      left panels, but zoomed into the respective bulk region.}
    \label{fig:obs}
\end{figure}

In Fig.~\ref{fig:obs} we illustrate the size of the off-shell effects
for a selection of kinematic distributions of final-state leptons and
$b$-quarks.  These distributions are not meant to compare realistic
predictions with different treatment of off-shell effects, but rather
our two datasets, so no event selection criteria are applied.  For the
invariant mass of the lepton-$b$ system we ensure, through charge
identification, that the two particles come from the same (anti)top
decay. Correspondingly, their invariant mass has an upper edge that
does not exist for off-shell events.  For the reconstructed top mass,
or the invariant mass of the three decay products, we clearly see the
Breit-Wigner propagator form, with an explicit cutoff.  Far below the
actual top mass, the off-shell prediction develops a shoulder at the
$W$-mass.  The secondary panels show the ratios of the integrated
one-dimensional phase space densities, illustrating that a reweighting
strategy between the two samples is unlikely to work.\medskip

%----------------------------------------------------------
%----------------------------------------------------------

The strategy behind our surrogate network is to generate an off-shell
event dataset and learn its structures relative to a corresponding
on-shell event dataset.  This strategy can be applied to any process
and at any order in perturbation theory.  Our two datasets consist of 5M
unit-weight events each. The six particles in the final state
of Eq.\eqref{eq:proc} are represented by $\{p_T, \eta, \phi \}$, with
fixed external particle masses. We remove three degrees of
freedom through a global azimuthal rotation and by enforcing
two-dimensional transverse momentum conservation, leaving us with a
15-dimensional phase space. Each on-shell requirement replaces a full
phase space dimension by a fixed range, given by the Breit-Wigner
shape with a hard-coded cutoff in the reconstructed invariant
mass.
Moreover, we conceal information relevant for eventual parton showering like the colourflow configuration or resonance history assignment. This is appropriate because the \textsc{Bb4l} generator in its default setup does not distinguish between the $t\bar{t}$ and single-top resonance histories and assigns colourflow correspondingly.

For the network input, the kinematic variables are preprocessed: we scale the transverse
momenta to $p_\mathrm{T}^{1/3}$ and express the azimuthal angles as 
$\mathrm{arctanh}(\phi/\pi)$. As subsequent classifier input, it turns out that 
$p_\mathrm{T}^{-1/3}$ leads to the best results. Finally, all input dimensions are standardized to zero mean and unit variance. 

%%%%%%%%%%%%%%%%%%%%%%%%%%%%%%%%%%%%%%%%%%%%%%%%%%%%%%%%%%%%%%%%%%%%%%%%
\section{Direct Diffusion}
\label{sec:diff}

For the network training, we start with the condition that we do not
want to train on paired on-shell and off-shell events, because such a
pairing does not follow a well-defined algorithm. This does not mean
it is impossible to construct such a mapping, but we expect it to lead
to artifacts.  Instead, we will train a generative Direct Diffusion (DiDi) network
on two phase space densities, one from on-shell and one from off-shell
events.

Usually, we employ generative networks to learn a mapping between a
simple latent space and some kind of phase space. The distribution in
the latent space is sampled from, so we typically use a uniform or a
Gaussian distribution.  In this application, the latent space
corresponds to the on-shell phase space, and the mapping is trained to
generate off-shell events,
\begin{align}
  x \sim p_{\text{on}}(x)
\quad 
\xleftrightarrow{\hspace*{1.5cm}}
\quad 
  x \sim \pmd(x|\theta) \approx p_\text{off}(x)  \; .
\label{eq:generative}
\end{align}
Using the conditional flow matching (CFM) setup for LHC
events~\cite{FM2022,Butter:2023fov}, we encode the transformation from on- to
off-shell events as a continuous time evolution, which follows an
ordinary differential equation (ODE)
\begin{align}
    \frac{dx(t)}{dt} \equiv v(x(t),t) \; .
\label{eq:sample_ODE}
\end{align}
In terms of the related probability density, we can formulate the same
task as
\begin{align}
\frac{\partial p(x,t)}{\partial t} + \nabla_x \left[ p(x,t) v(x,t) \right] = 0 \; .
\label{eq:continuity}
\end{align} 
The ODE and the continuity equation are equivalent, so we can train a
network to represent the velocity field and use this velocity field to
generate samples using a fast ODE solver for Eq.\eqref{eq:sample_ODE}. Defining our time dependent probability density as
\begin{align}
 p(x,t) \to 
 \begin{cases}
  p_\text{off}(x) \quad & t \to 0 \\
  p_\text{on}(x)  \quad & t \to 1  \;,
\end{cases} 
\label{eq:fm_limits}
\end{align}
we now need to construct the associated velocity field. 

%----------------------------------------------------------
\begin{figure}[t]
    \centering
    \includegraphics[width=0.75\textwidth]{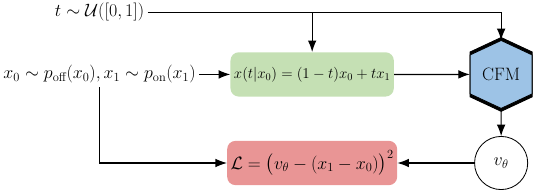}
    \caption{Training procedure for a CFM network mapping between on-shell and off-shell phase space distributions. Diagram adapted from Ref~\cite{Butter:2023fov}.}
    \label{fig:training_setup}
\end{figure}
%---------------------------------------------------------

Unlike for the usual generative setup, we define $x(1) = x_1$ as a
sample from the on-shell phase space, whereas $x(0)=x_0$ corresponds to the off-shell phase space. 
Therefore, we adapt the linear trajectory between on-shell and off-shell
events to
\begin{align}
    x(t|x_0) 
%    = \beta_t x_0 + \sigma_t\epsilon 
    = (1-t) x_0 + t x_1
\to \begin{cases}
      x_0 \quad & t \to 0 \\
      x_1 \sim p_\text{on} \quad & t \to 1 \;.
\end{cases} 
\label{eq:gaussian_probability_path_reparametrization}
\end{align}
The true conditional velocity field of our linear trajectory linked to our probability density in Eq.~\eqref{eq:fm_limits} is hence given by
\begin{align}
   v(x(t|x_0),t| x_0) 
%   &= \frac{d x(t|x_0)}{dt}\notag \\
   &= \frac{d}{dt} \left[ (1-t) x_0 + t x_1 \right] = - x_0 + x_1 \; .
\label{eq:conditional_velocity}
\end{align}
The remaining derivation is in complete analogy to
Ref.~\cite{Butter:2023fov}, leading to the simple MSE loss
\begin{align}
    \mathcal{L}_\text{CFM} &= \bigl\langle \left[ v_\theta((1-t)x_0+t x_1,t) - (x_1 - x_0)\right]^2 \bigr\rangle_{t \sim \mathcal{U}([0,1]),x_0\sim p_\text{off}, x_1 \sim p_\text{on}}  \; .
\label{eq:CFM_loss}
\end{align}
Because the CFM setup does not include a likelihood, we can go
directly from $p_\text{on}$ to $p_\text{off}$, without extra effort
due to detours like in Flows4Flows~\cite{Golling:2023mqx}, and without
any pairing between $x_0 \sim p_\text{off}$ and $x_1 \sim
p_\text{on}$.\medskip

As usual, we use a Bayesian version of the generative network~\cite{Bellagente:2021yyh,Butter:2021csz}, to extract 
uncertainties on the learned phase space density.
The Bayesian CFM loss~\cite{Butter:2023fov} 
\begin{align}
    \loss_\text{B-CFM} &= \XLangle \loss_\text{CFM}\XRangle_{\theta \sim q(\theta)} + c  \kl[q(\theta),p(\theta)]
%    &= \XLangle \big( v_t^\theta(x_t) - v_t(x_t;x_0,\epsilon) \big)^2\XRangle_{t\sim U([0,1]),x_0\sim \pd(x_0), \epsilon \sim \normal(0,1)}......\notag \\
\label{eq:BCFM_objective}
\end{align}
includes a hyperparameter $c$ to balance the regular CFM loss with the Bayesian 
regularization term. If the first loss term was a likelihood loss, this factor 
would be fixed by Bayes' theorem. We have checked that the network performance is stable over many orders of magnitudes for $c = 10^{-10}~...~10^{-2}$. For larger values we observe that the training becomes unstable, as expected, while for very small values the uncertainty can no longer be captured. In addition, the 
prior weight  distribution $q(\theta)$ is given by a unit Gaussian, where the choice of 
width hardly affects the network performance. Compared to the  
deterministic counterpart, this Bayesian network is equally precise.

%----------------------------------------------------------
\begin{table}[b!]
\centering
\begin{small} \begin{tabular}{l|c}
\toprule
     Hyperparameter               \\
     \midrule
     Embedding dimension      & 64 \\
     Layers                   & 8  \\
     Intermediate dimensions  &768 \\
     \midrule
     LR scheduling           & OneCycle \\ 
     Starter LR              & $10^{-4}$ \\ 
     Max LR                  & $10^{-3}$ \\
     Epochs                  & 1000 \\
     Batch size              & 16384 \\
     \midrule
     $c$                        & $10^{-3}$\\
     \midrule
     \# Training events      & 3 M \\
     \bottomrule
\end{tabular} \end{small}
\caption{Generative network setup (DiDi) and hyperparameters.}
\label{tab:hyper}
\end{table}
%----------------------------------------------------------

The network and training setup is visualized in
Fig.~\ref{fig:training_setup}. We use a simple dense network with SiLU activation, 
where the last layer is initialized at zero. This sets the initial velocity 
field to zero and induces an identity mapping at the starting point of the training. 
We do not enforce a prescription for
turning a given on-shell event into an off-shell event
through the training data or the training procedure. 
In fact, for each epoch 
different phase space points will be connected 
via a linear trajectory. Instead, the network training constructs its
own mapping to populate the off-shell phase space. This happens as part of the 
loss minimization, which means the transport map follows from an implicit measure encoded 
in the network architecture and loss.\medskip

%----------------------------------------------------------
\begin{figure}[t]
  \includegraphics[width=0.49\textwidth, page=6]{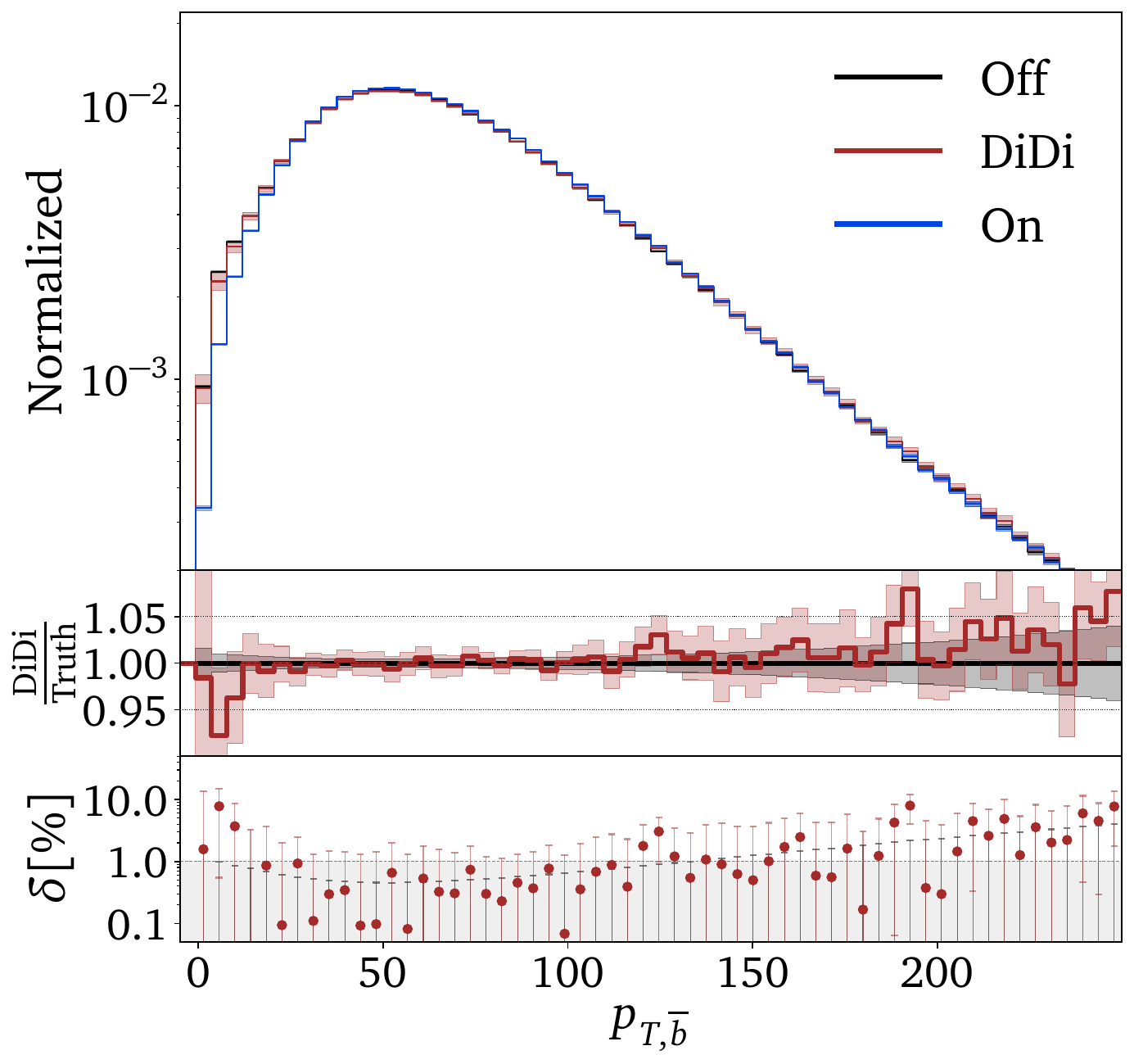}
  \includegraphics[width=0.49\textwidth, page=19]{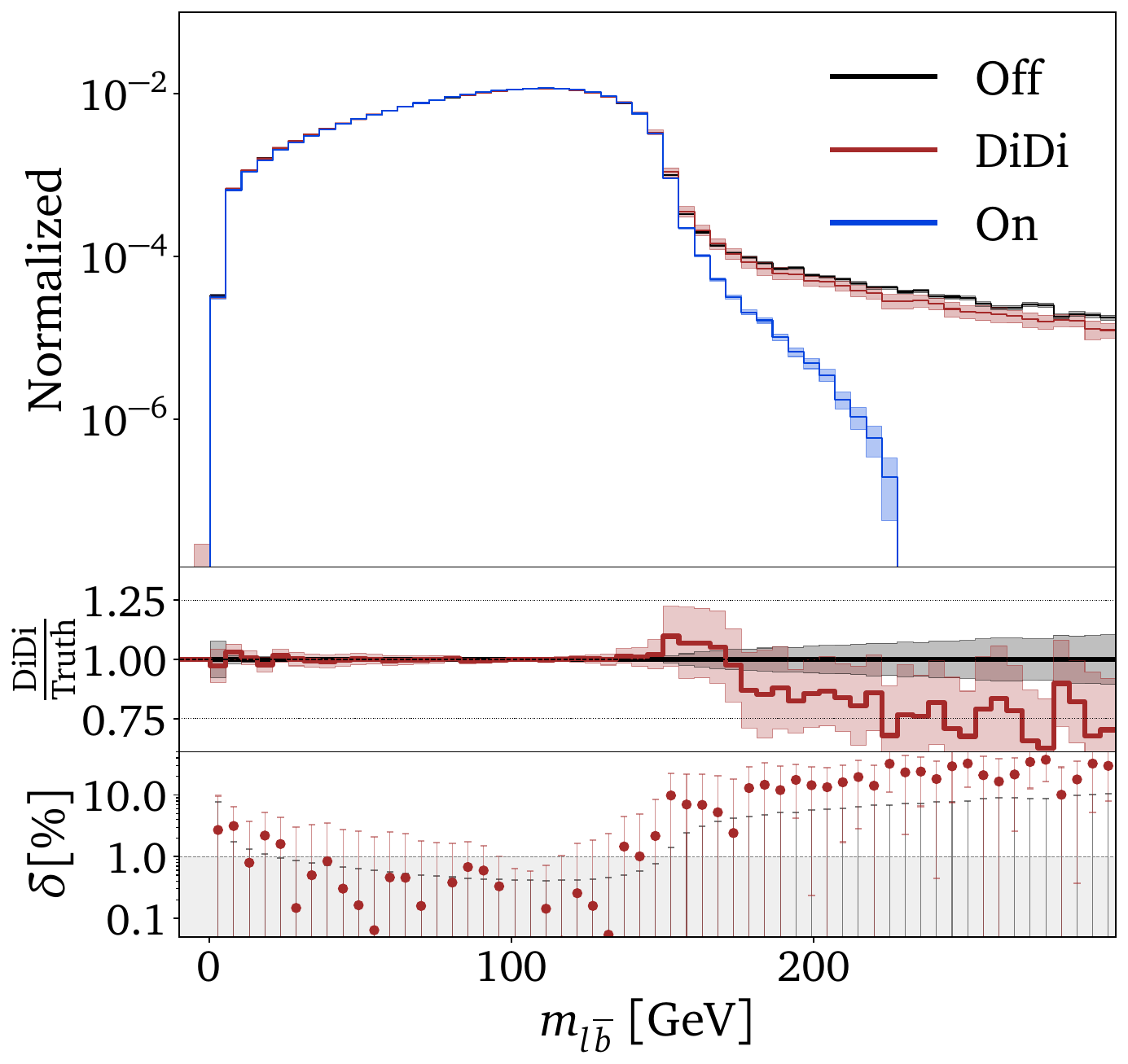}\\
  \includegraphics[width=0.49\textwidth, page=88]{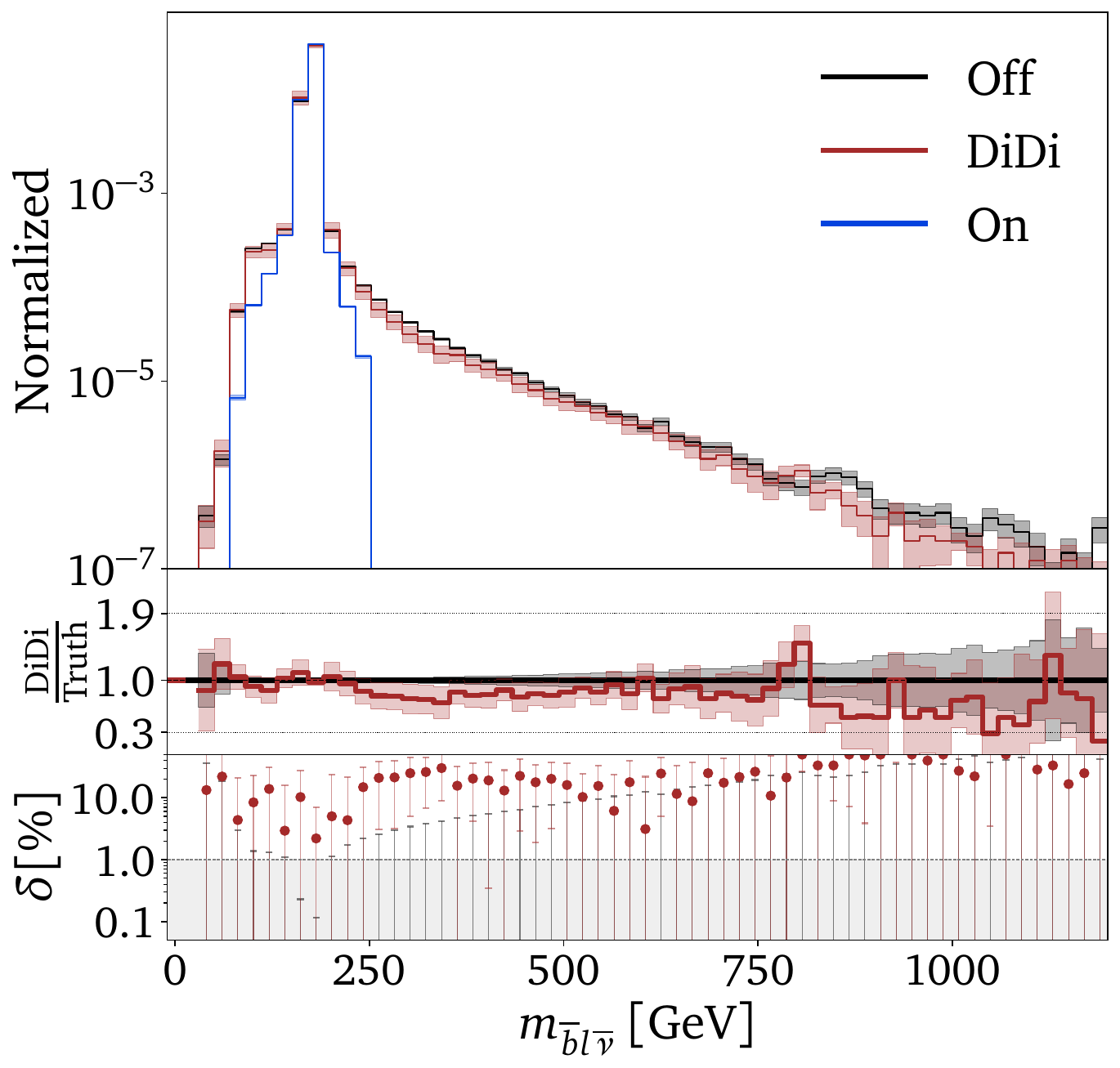}    
  \includegraphics[width=0.49\textwidth, page=17]{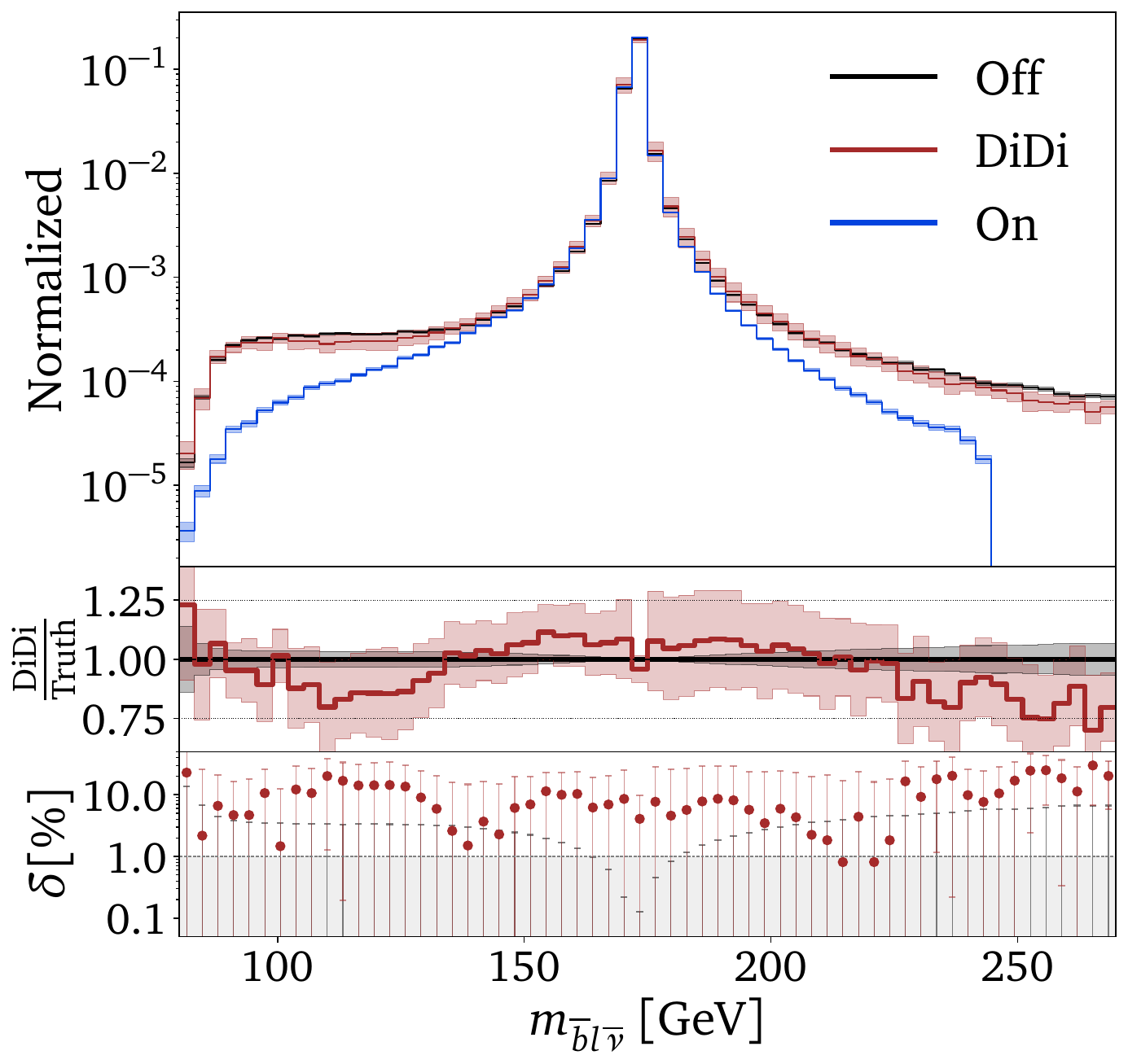}
  \caption{Results from our Direct Diffusion off-shell generator,
    compared to the on-shell starting point and the off-shell training
    distributions.}
  \label{fig:D2F}
\end{figure}
%----------------------------------------------------------

As the CFM training objective is a simple regression, we can train on 17 dimensions,
including two redundant degrees of freedom, as this happens to increase the precision.
The two additional observables, the transverse momentum and the polar angle of the neutrino, are determined by transverse momentum conservation and are hence multidimensional correlations. Empirically we found that it is easier for the model to learn the behavior of those observables when handed directly. Especially more complicated correlations such as the invariant mass of the reconstructed top benefit greatly from this additional information. While improving the efficiency of the training these dimensions will be ignored for the actual event generation.

The network hyperparameters and the training parameters are 
given in Tab.~\ref{tab:hyper}. 
We encode $t$ in a higher embedding dimension and following~\cite{FM2022} we add batch-wise random noise of scale $10^{-4}$ to $x_0$ and $x_1$ during training. We use the standard \textit{Dopri5} ODE solver 
to sample from our network.
In the interest of precision we use a large batch size. This is  
a problem for generic optimal-transport networks~\cite{OTCFM}, 
but can be easily implemented in our architecture. We tested 
the OT-CFM~\cite{OTCFM} and found that in our setup the required small batch size indeed led to worse performance.\medskip

In Fig.~\ref{fig:D2F} we show a set of one-dimensional kinematic
distributions, for the on-shell data we start from, the off-shell training data, and the generated off-shell data. In the first panel we see how the network learns subtle
differences almost perfectly well. The typical precision is around the
per-cent level.  For complex and sensitive correlations, like the
lepton-$b$ invariant mass and the reconstructed top mass we start with
a huge deviation between the on-shell distribution and the off-shell
target.  To generate these distributions correctly, the generative network
has to learn correlations in the corresponding 9-dimensional sub phase space.

%----------------------------------------------------------
\begin{figure}[t]
    \includegraphics[page=4, width=0.495\textwidth]{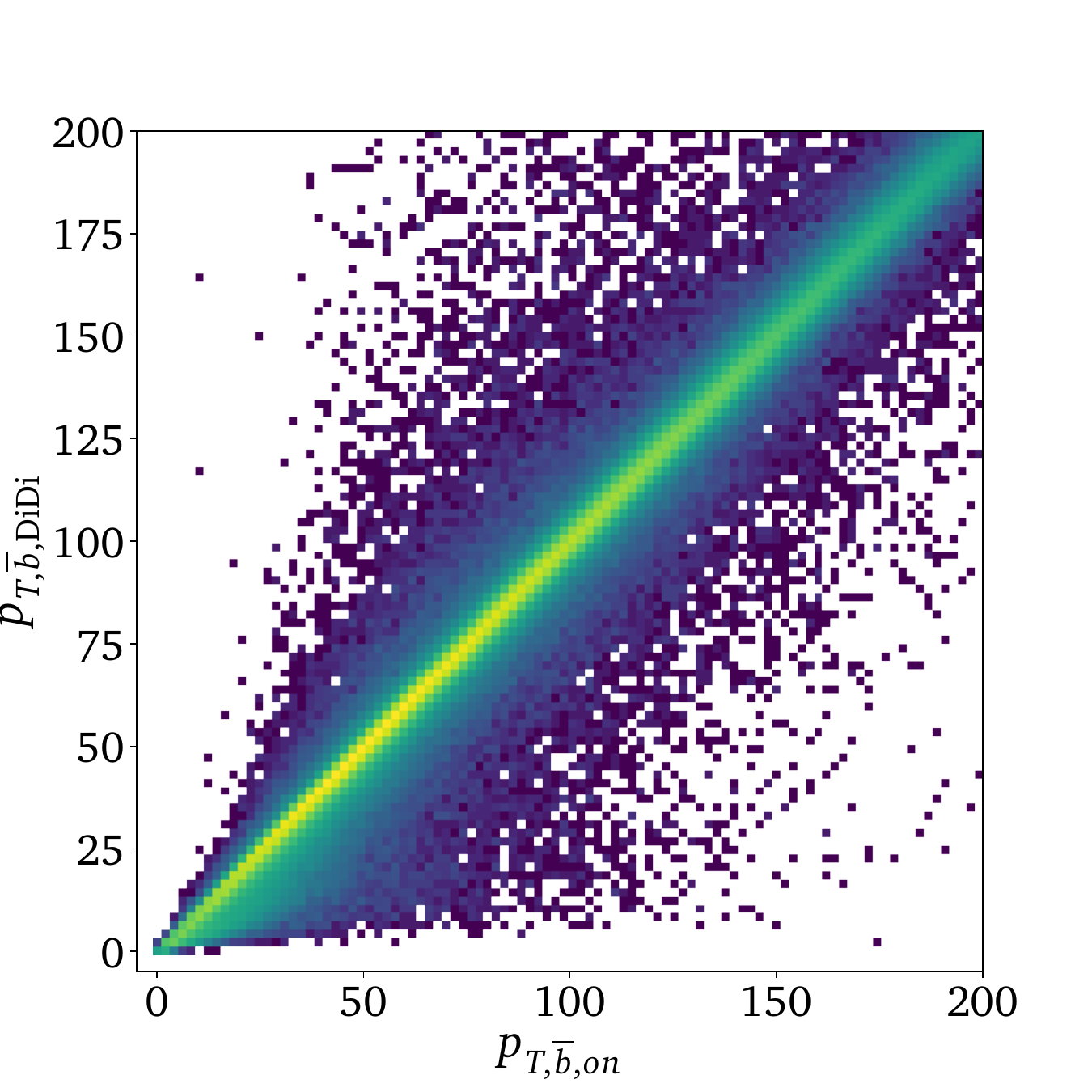}
    \includegraphics[page=2, width=0.495\textwidth]{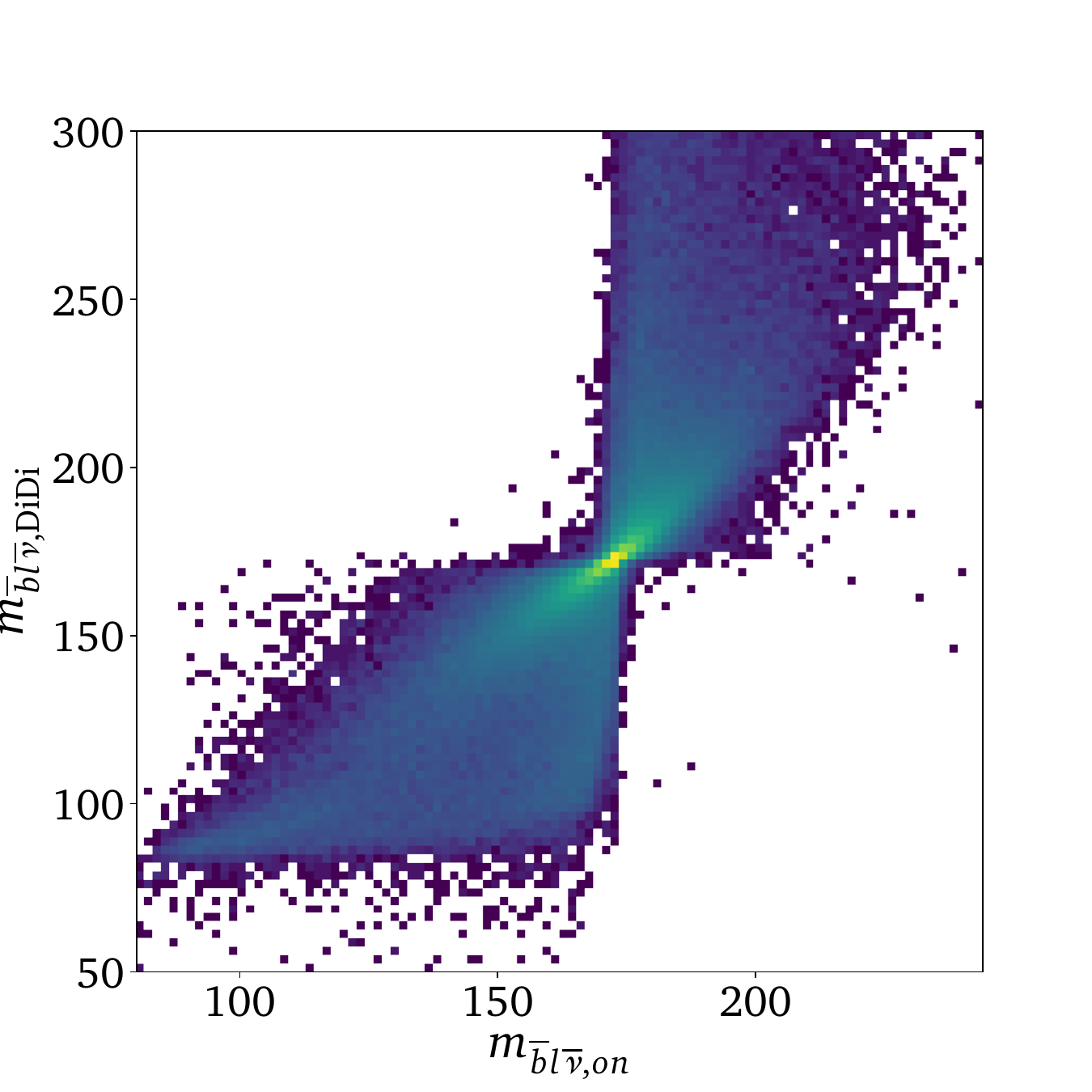}
    \caption{Migration plot --- correlation between generated
      (off-shell) and starting (on-shell) distributions for two
      kinematic correlations. It illustrates the mapping found by the network during training.}
    \label{fig:migration}
\end{figure}
%----------------------------------------------------------

Our results confirm that in phase space regions covered well by both
datasets, the generative network reproduces the target distribution
precisely, as one could expect. However, we also see that even in
phase space regions not populated by on-shell events the target
distribution is reproduced relatively precisely. The typical agreement
between the generated and target densities is around 10\% in phase
space regions with relatively little training data. While this deviation is covered by the uncertainties of the Bayesian network, we propose 
possible improvements in the next section.\medskip

Finally, we can ask how our generative network fills the off-shell
phase space from the on-shell events.  In Fig.~\ref{fig:migration} we
show the correlations between the generated off-shell distribution and
the on-shell starting distributions, \ie the migration of paired
latent and target phase space events from the forward simulation,
where we emphasize that the pairing is only defined by the network
evaluation, not by the training. In general, the correlation between
the kinematic observables should be close to the identity, as confirmed by the $p_{T,\bar{b}}$
distribution in the left panel. This means that the shift from on-shell to off-shell
phase space is relatively small and uncorrelated. However, for the
reconstructed top mass, some of the events have to be shifted by a
larger value, as illustrated in the right panel.  While the width of
the linear correlation becomes very small around the top mass peak, it
rapidly increases away from the peak, demonstrating the large shift required to populate the off-shell phase space. Moreover, our network maps
events from each side of the Breit-Wigner peak to the same side of the
off-shell distribution.

%%%%%%%%%%%%%%%%%%%%%%%%%%%%%%%%%%%%%%%%%%%%%%%%%%%%%%%%%%%%%%%%%%%%%%%%
\section{Classifier control and reweighting}
\label{sec:reweight}

Because the unsupervised density estimation underlying our generative
network is more challenging and less precise than a
supervised classifier training, we can use a trained classifier as a
function of phase space to systematically check and improve the
generative network.  A perfectly trained statistical classifier
converges to the ratio of likelihoods,
\begin{align}
    C(x) = \frac{p_\text{data}(x)}{p_\text{data}(x)+p_\text{model}(x)} \; .
\end{align}
As a function of phase space we can use this classifier to construct an
event reweighting, which improves the precision of the generative
networks~\cite{Butter:2021csz},
\begin{align}
    w(x) = \frac{p_\text{data}(x)}{p_\text{model}(x)} = \frac{C(x)}{1-C(x)} \; .
\label{eq:weights}
\end{align}
In addition, we can use the same learned event weights to determine
the precision of the generative network and systematically search for
failure modes by searching for clusters of very small or very large
weights in phase space~\cite{Das:2023ktd}.

%----------------------------------------------------------
\begin{figure}[t]
  \includegraphics[width=0.49\textwidth,page=6]{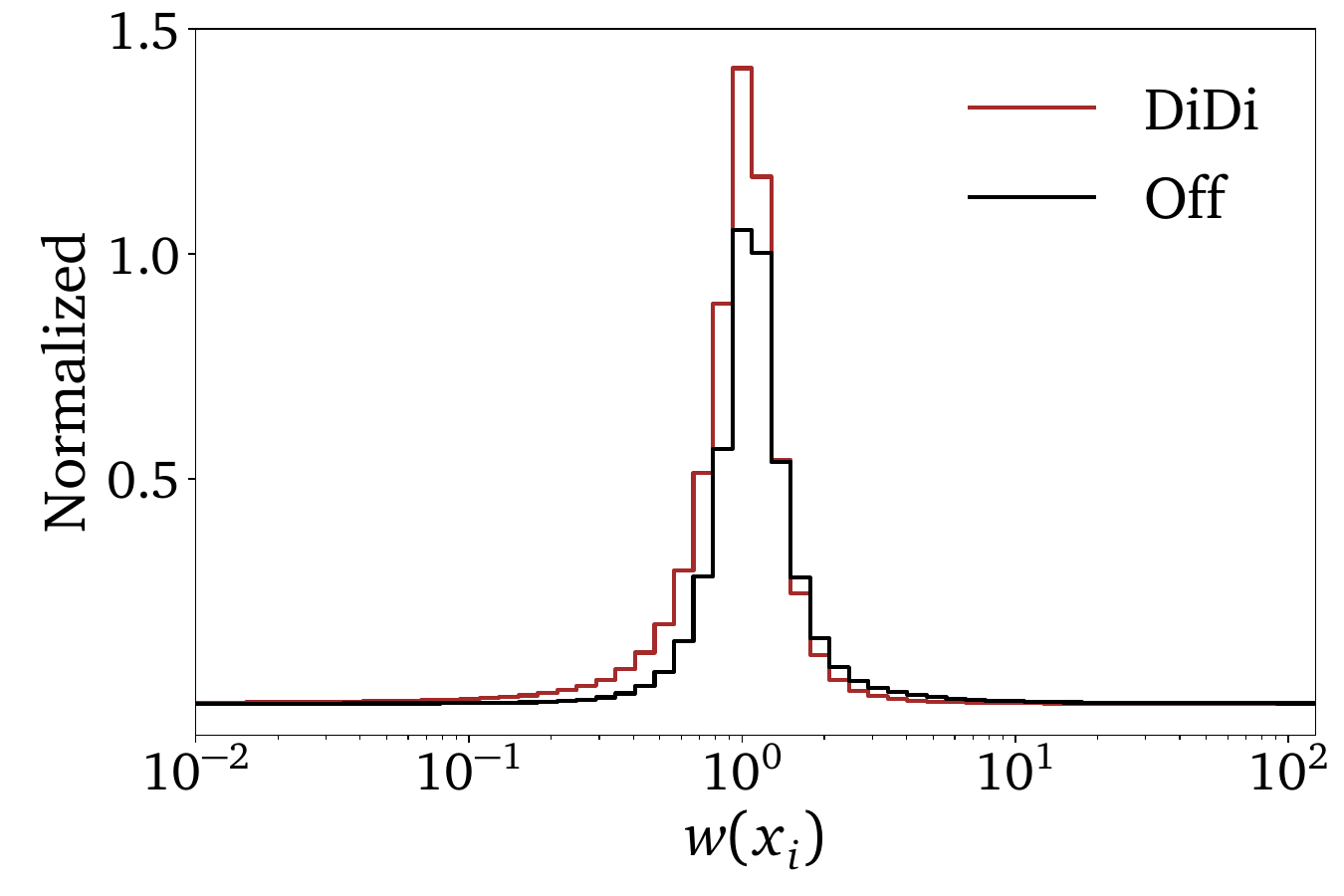}
  \includegraphics[width=0.49\textwidth,page=4]{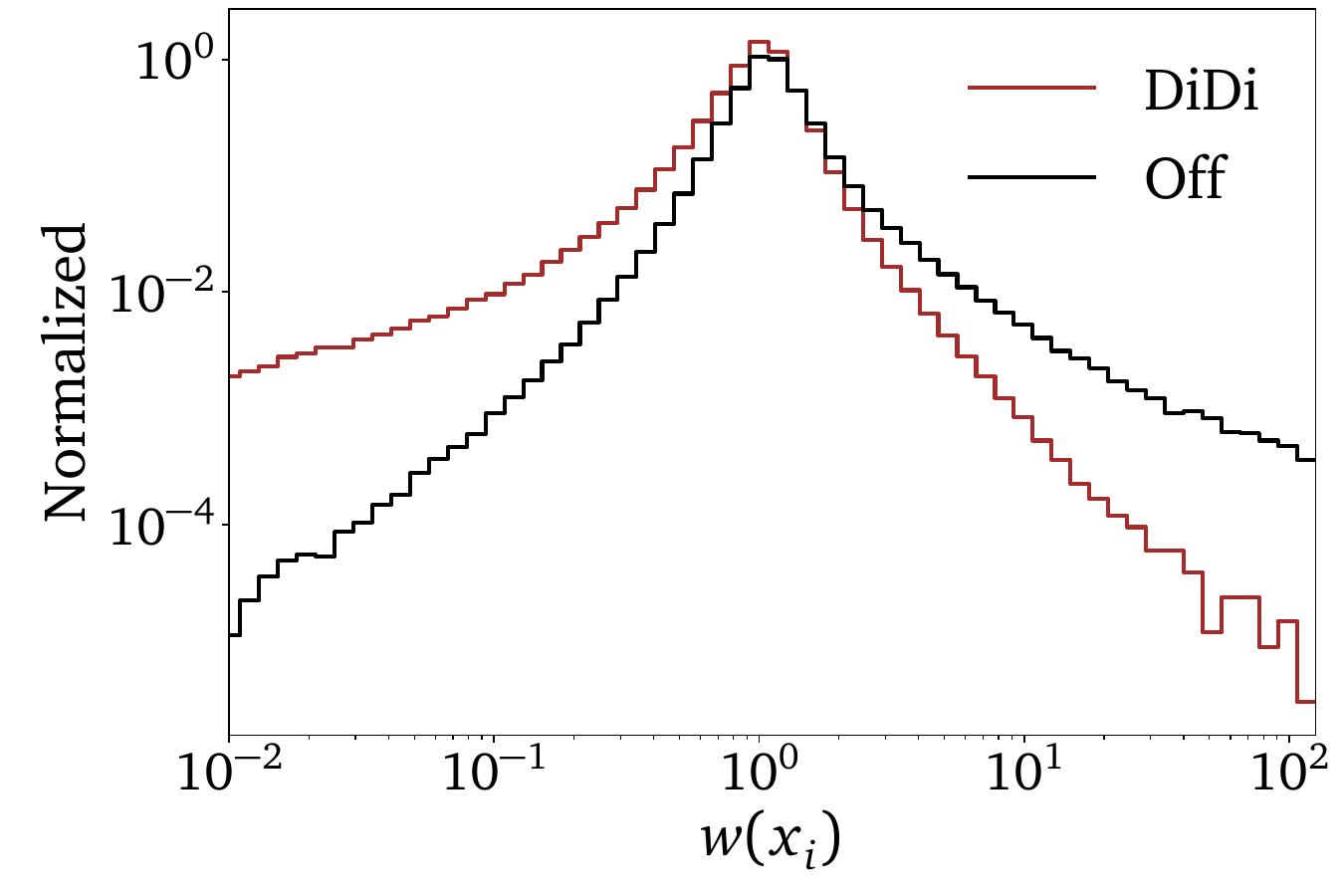}
  \caption{Histogram of the learned event weights, evaluated on the
    off-shell training data and the DiDi-generated
    off-shell events. The two panels show the same weights, on 
    linear and logarithmic scales.}
  \label{fig:histo}
\end{figure}
%----------------------------------------------------------

%----------------------------------------------------------
\begin{table}[b!]
\centering
\begin{small} \begin{tabular}{l|c}
\toprule
     Hyperparameter               \\
     \midrule
     Layers                       & 5  \\
     Intermediate dimensions       &512 \\
     Dropout                 &  0.1 \\
     Normalization           & BatchNorm1d \\
     \midrule
     LR scheduling           & ReduceOnPlateau \\ 
     Starter LR              & $1^{-3}$ \\ 
     Patience                & 10 \\
     Epochs                  & 100 \\
     Batch size              & 1024 \\
    \midrule
     \# Training events      & 2.5 M \\
     \bottomrule
\end{tabular} \end{small}
\caption{Classifier network setup and hyperparameters.}
\label{tab:hyper_classifier}
\end{table}
%----------------------------------------------------------

We train the classifier on 27 observables, our 15 physical dimensions,
complemented by the reconstructed top and anti-top masses, the 
reconstructed $W^+$ and $W^-$ masses, the reconstructed masses of the 
$\bar{b}l$ and $b\bar{l}$ systems, and the six corresponding 
transverse momenta. For the input of our classifier we sample events from our Bayesian generator setting each network weight to its mean value. The setup is given in 
Tab.~\ref{tab:hyper_classifier}.\medskip

We show histograms of learned phase space weights
$w(x_i)$ on a linear and a logarithmic axis in Fig.~\ref{fig:histo}.
As expected, the distributions peak at unit
weights, with a width around $0.3$. Following Eq.\eqref{eq:weights}, large weights correspond
to phase space regions where the generative network produces a too
small density of off-shell points; small weights mark phase space
regions where the generative network produces too many off-shell
events, compared to the training data. To study both tails of the weight
distribution, we evaluate the weights over the combination of
2M training and 2M generated events and confirm that both 
tails decrease rapidly. We 
eventually clip the event weights to $w< 15$ to improve the numerical behavior of our generation and avoid sparks in regions of 
low statistics.\medskip

\begin{figure}[t]
  \includegraphics[width=0.49\textwidth, page=6]{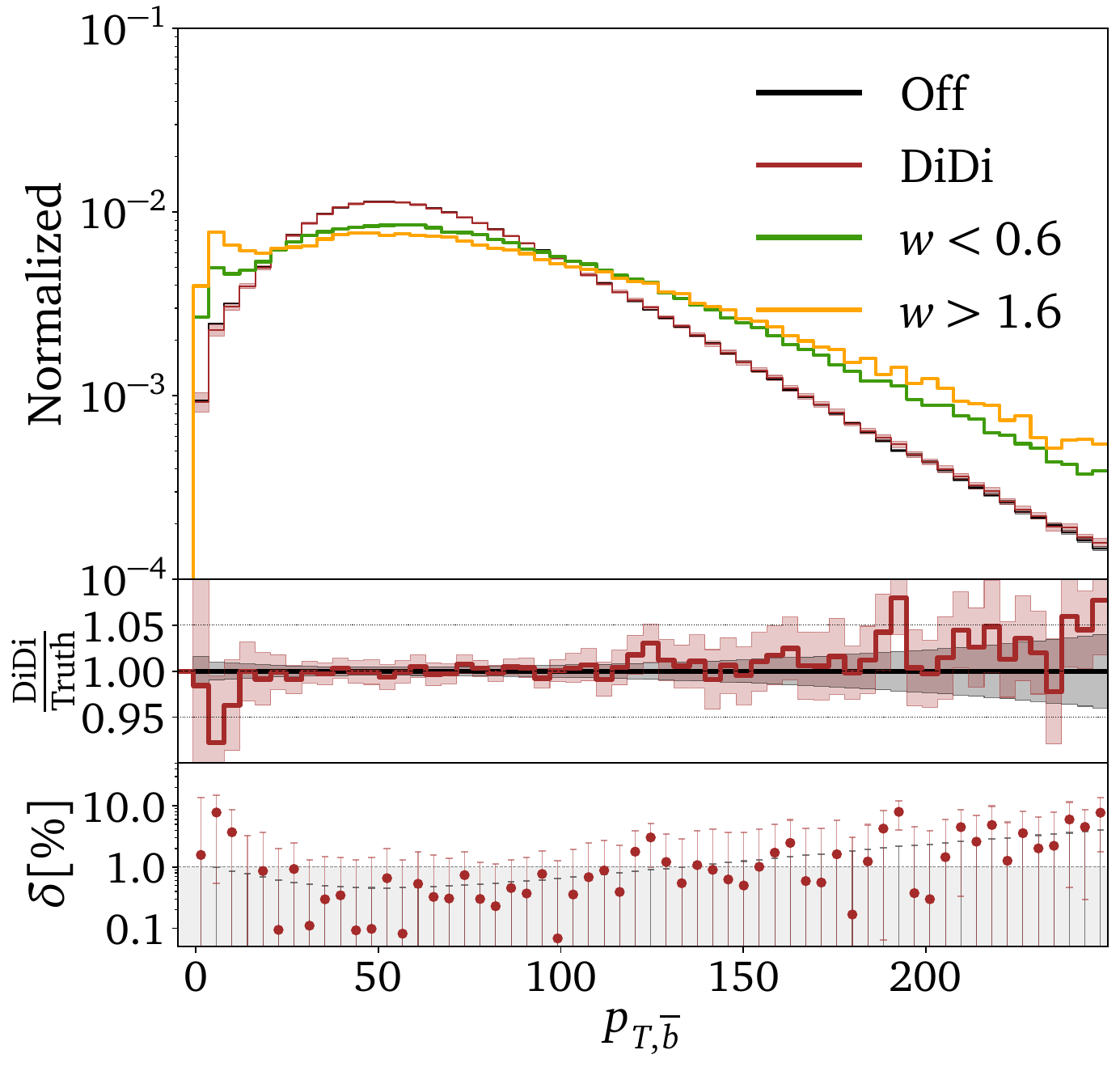}
  \includegraphics[width=0.49\textwidth, page=17]{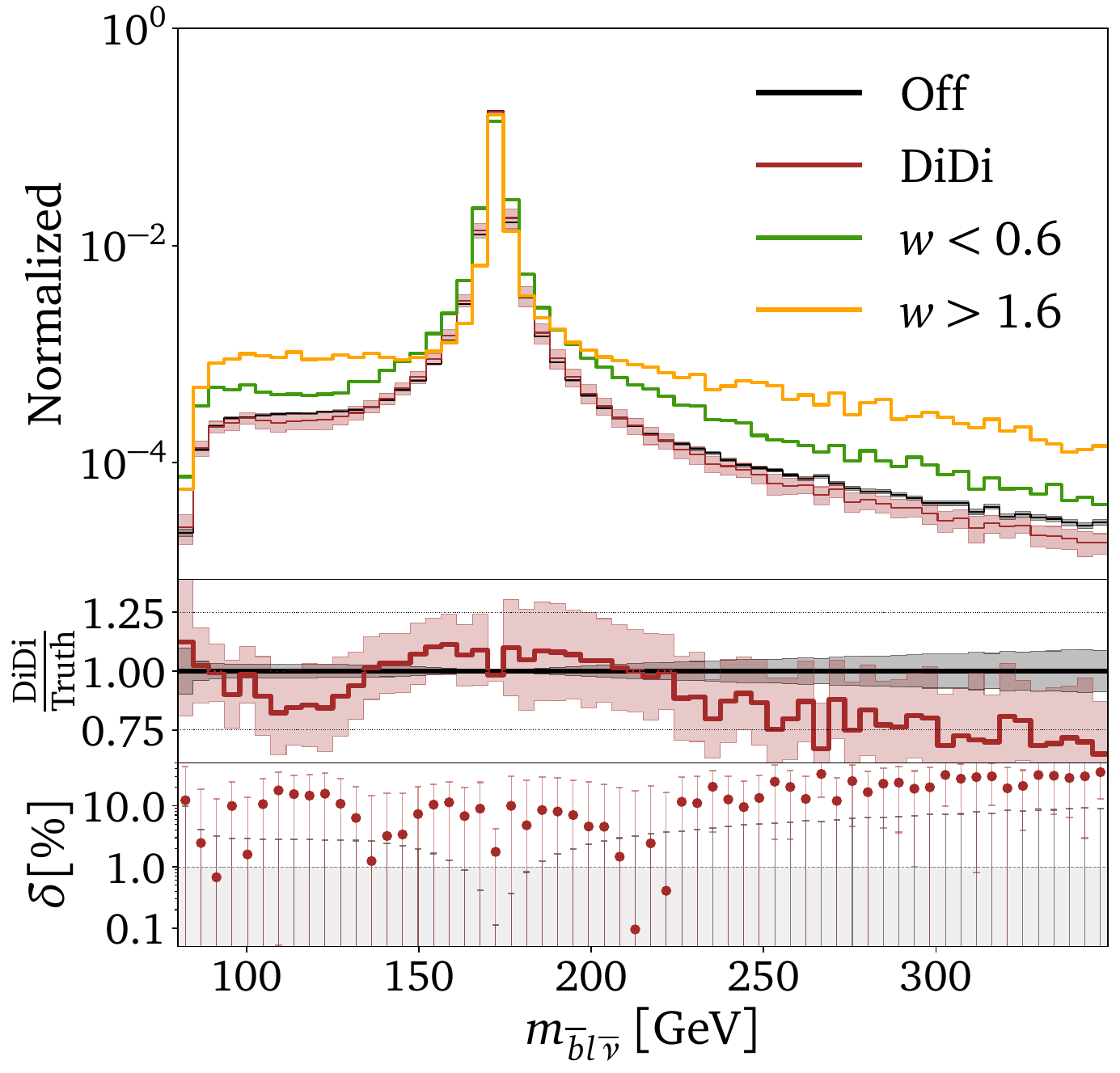}
   \caption{Clustering of the classifier trained to distinguish the
         off-shell training data from DiDi-generated events, for two example distributions.}
  \label{fig:clustering}
\end{figure}

In Fig.~\ref{fig:clustering} we track phase space regions with very
small and very large weights. We compare all events of our test sample to the subsets
with $w(x) < 0.6$, corresponding to 12.3\% of the sample, and
$w(x) > 1.6$, corresponding to 6.2\% of the generated
sample. The two shown distributions illustrate the general feature 
that the
kinematic distributions for both tails are similar. In the $p_{T,\bar{b}}$-distribution we can identify two 
limitations of the network training: for small transverse momenta
a slight shift of the cliff will lead to large relative weight
corrections, while for large transverse momenta the decreasing density
of training events will increase the relative size of the noise. For the 
reconstructed anti-top mass, one of the critical distributions for our 
generative task, the low-weight and high-weight tails are also comparable
with each other and also comparable on both sides of the mass peak. The only distinct feature appears around the peak where events with small weight dominate the slightly over populated sides of the peak.
This indicates that all weight tails are generated by noise, there are no
missing localized features, and the limiting factor is the statistics of 
the training data.\medskip

%----------------------------------------------------------
%----------------------------------------------------------

\begin{figure}[t!]
  \includegraphics[width=0.49\textwidth, page=6]{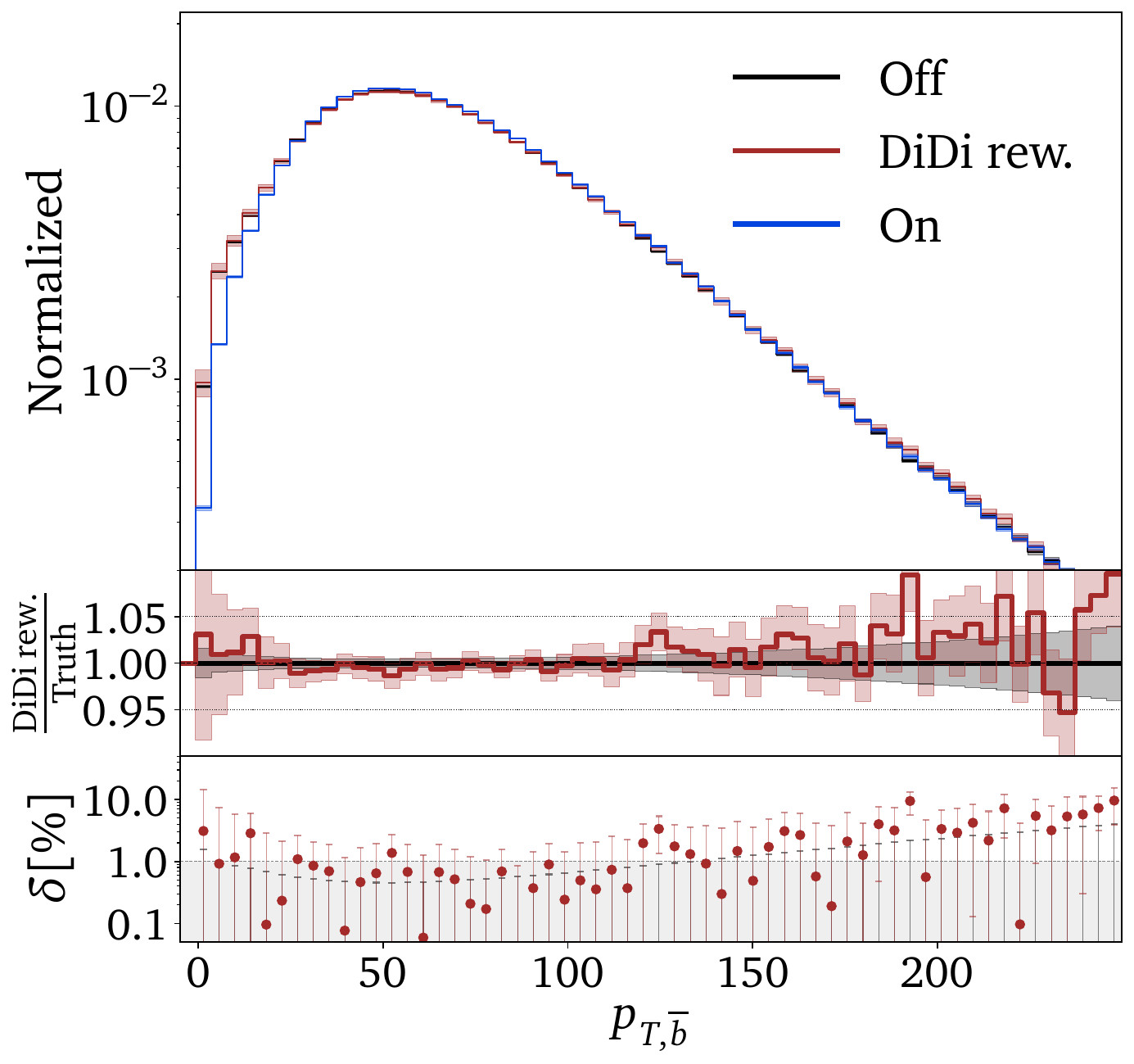}
  \includegraphics[width=0.49\textwidth, page=18]{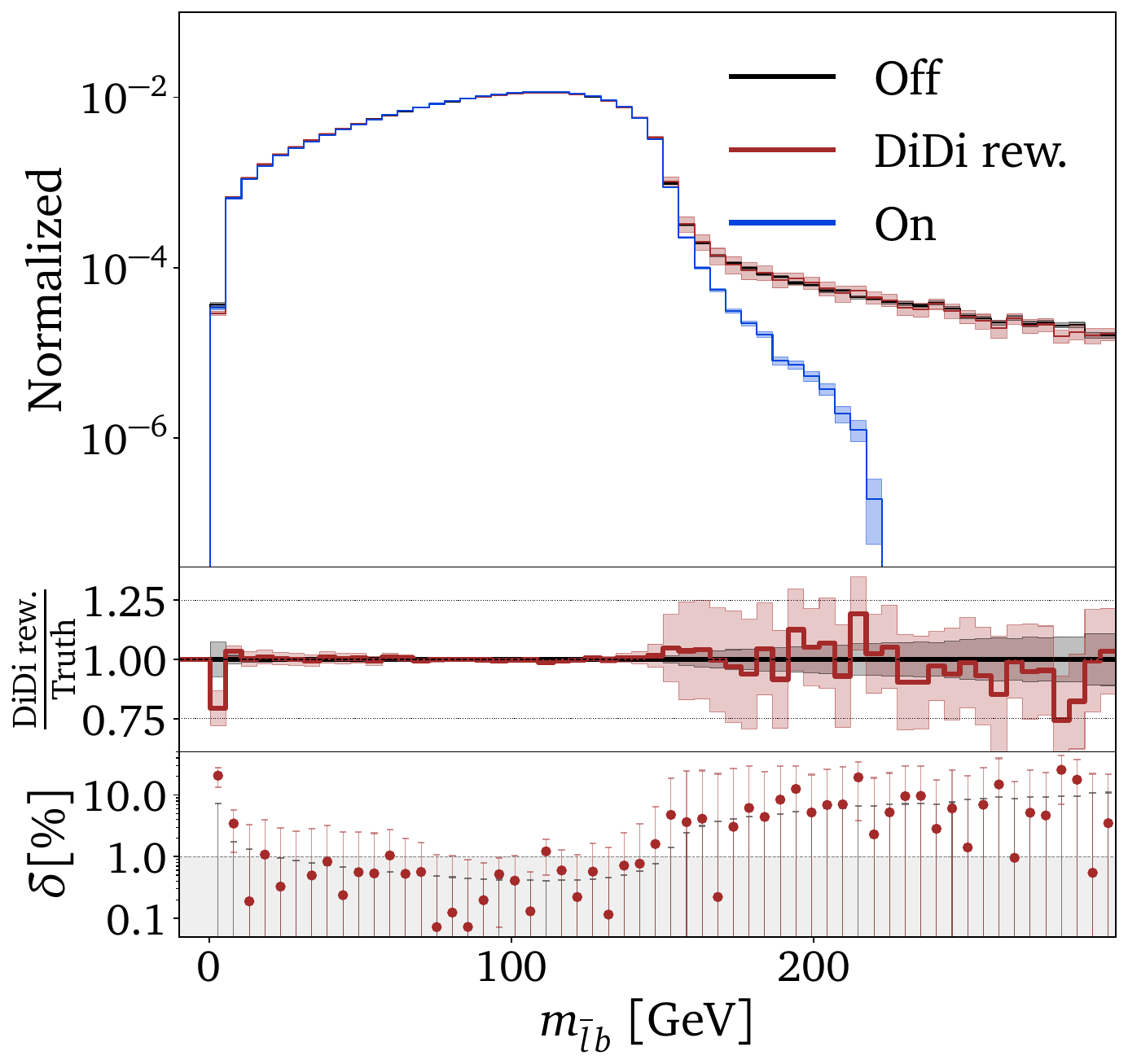}\\
  \includegraphics[width=0.49\textwidth, page=36]{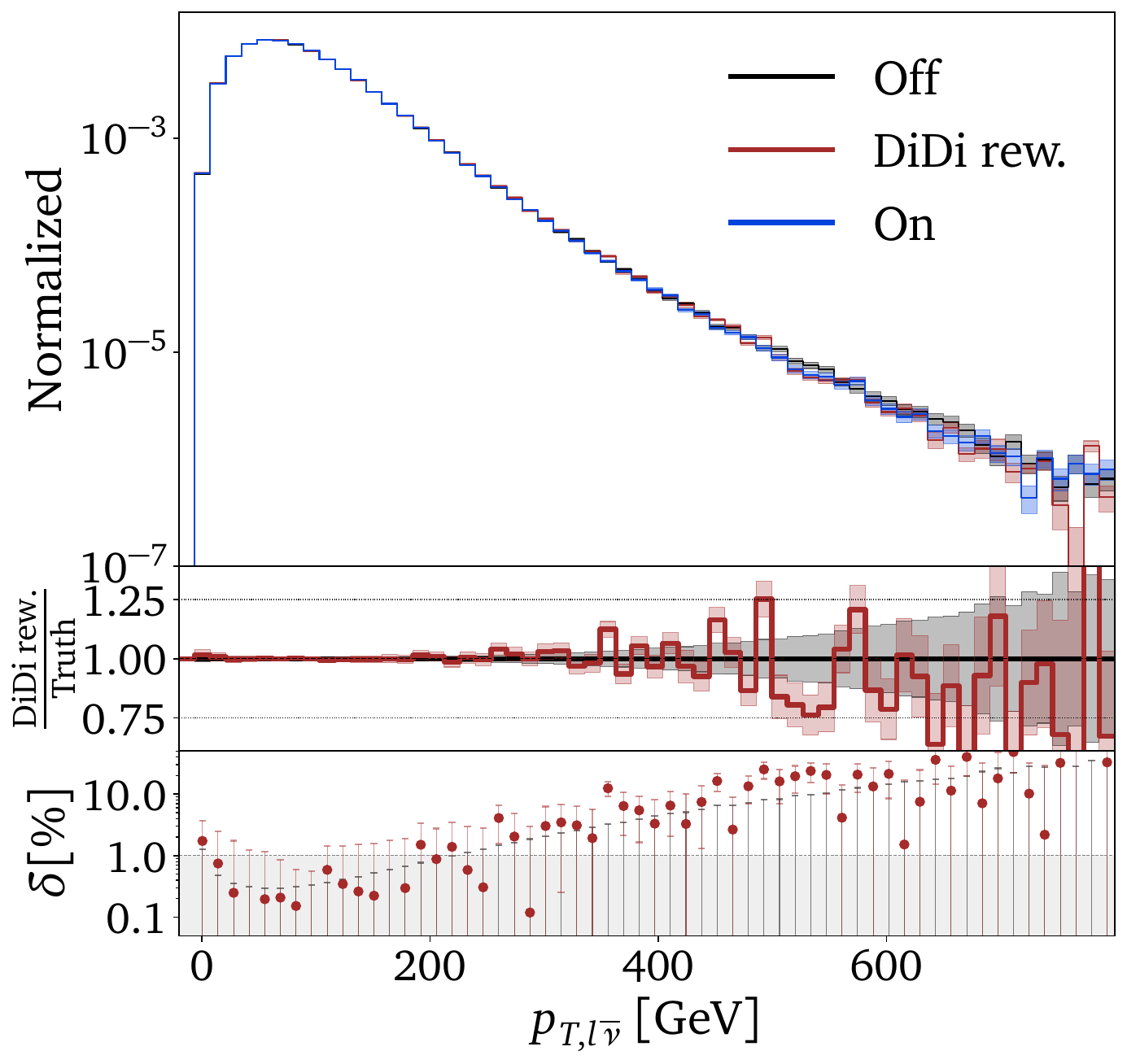}
  \includegraphics[width=0.49\textwidth, page=21]{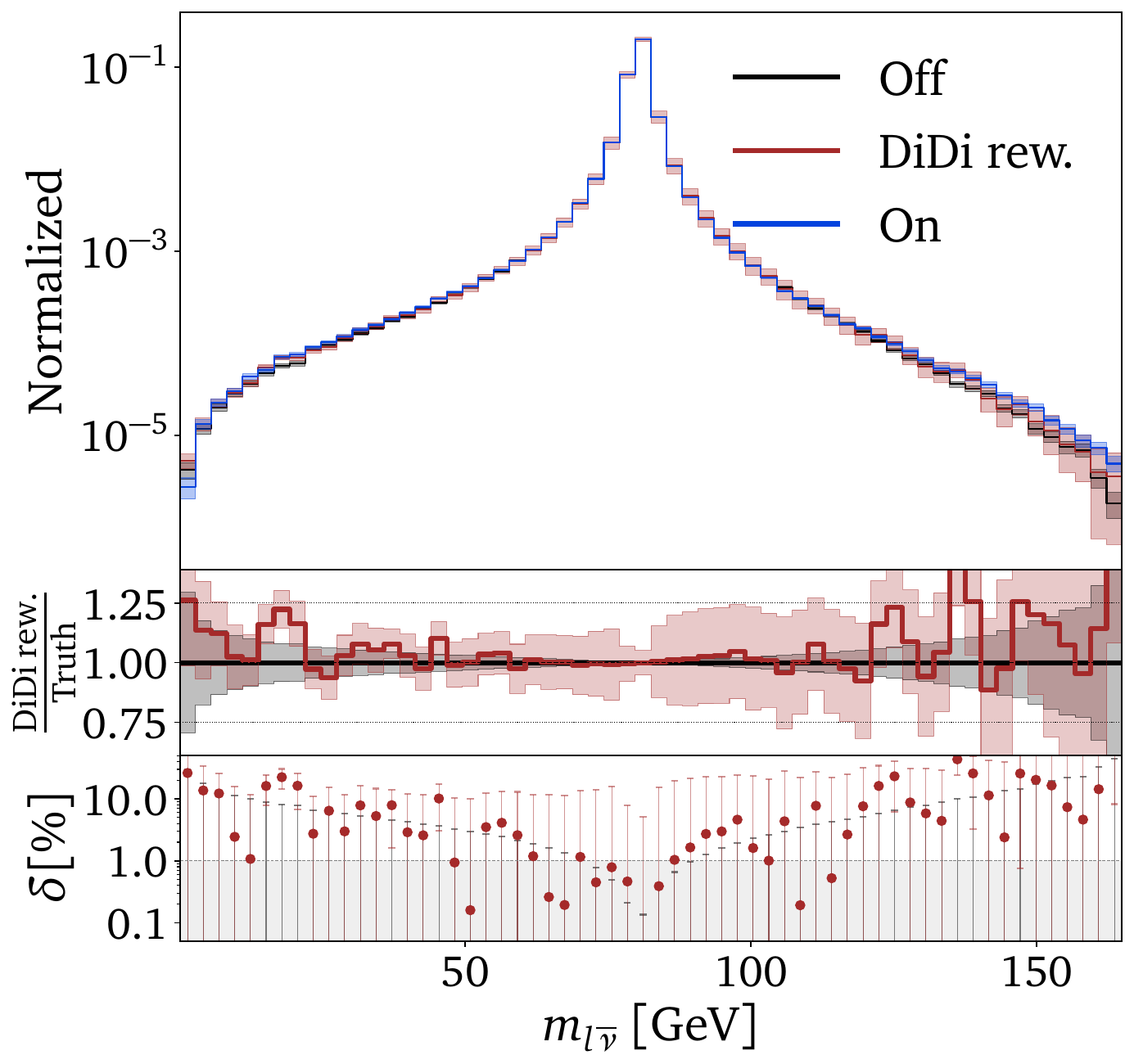}\\  
  \includegraphics[width=0.49\textwidth, page=88]{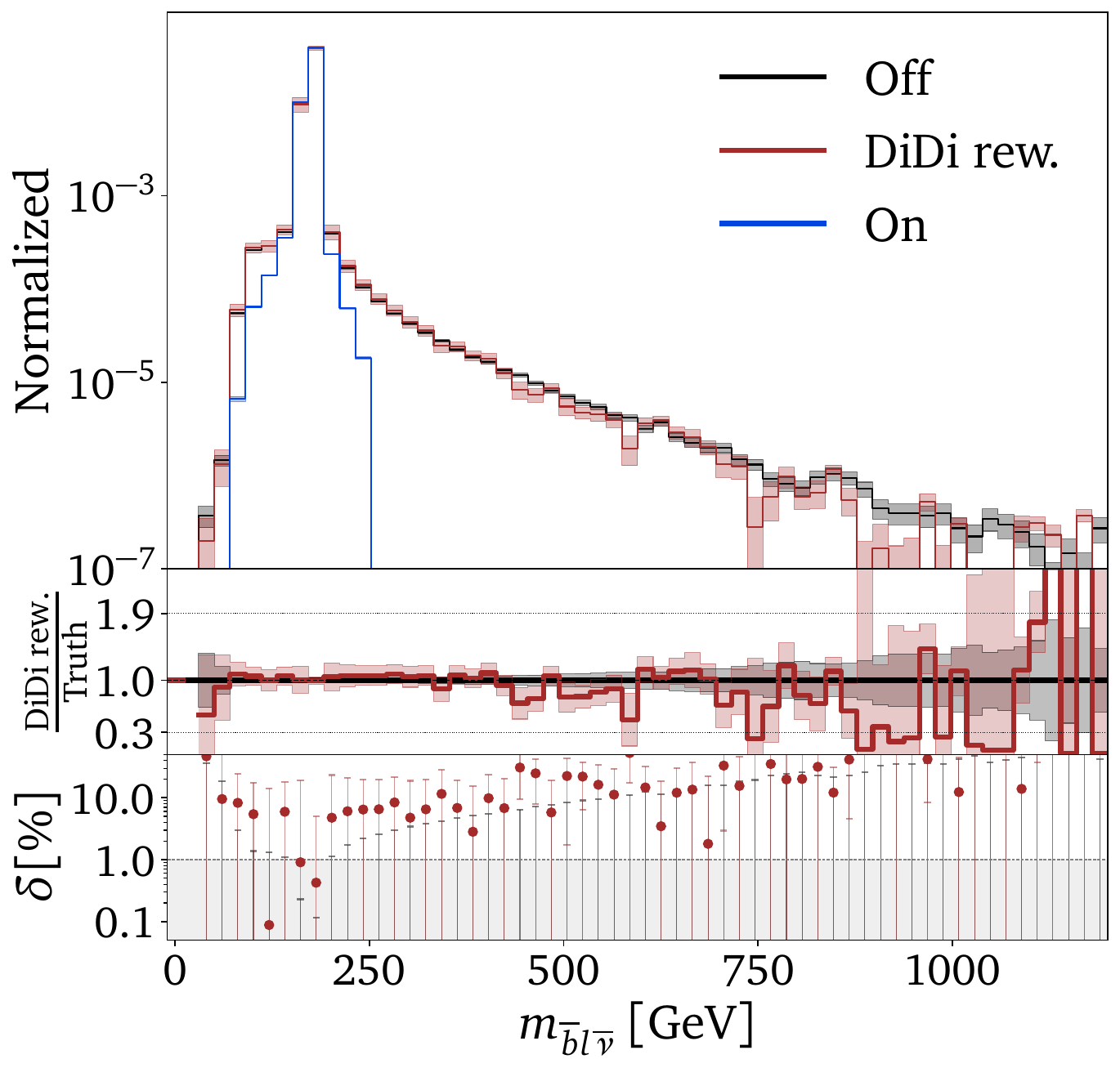}    
  \includegraphics[width=0.49\textwidth, page=17]{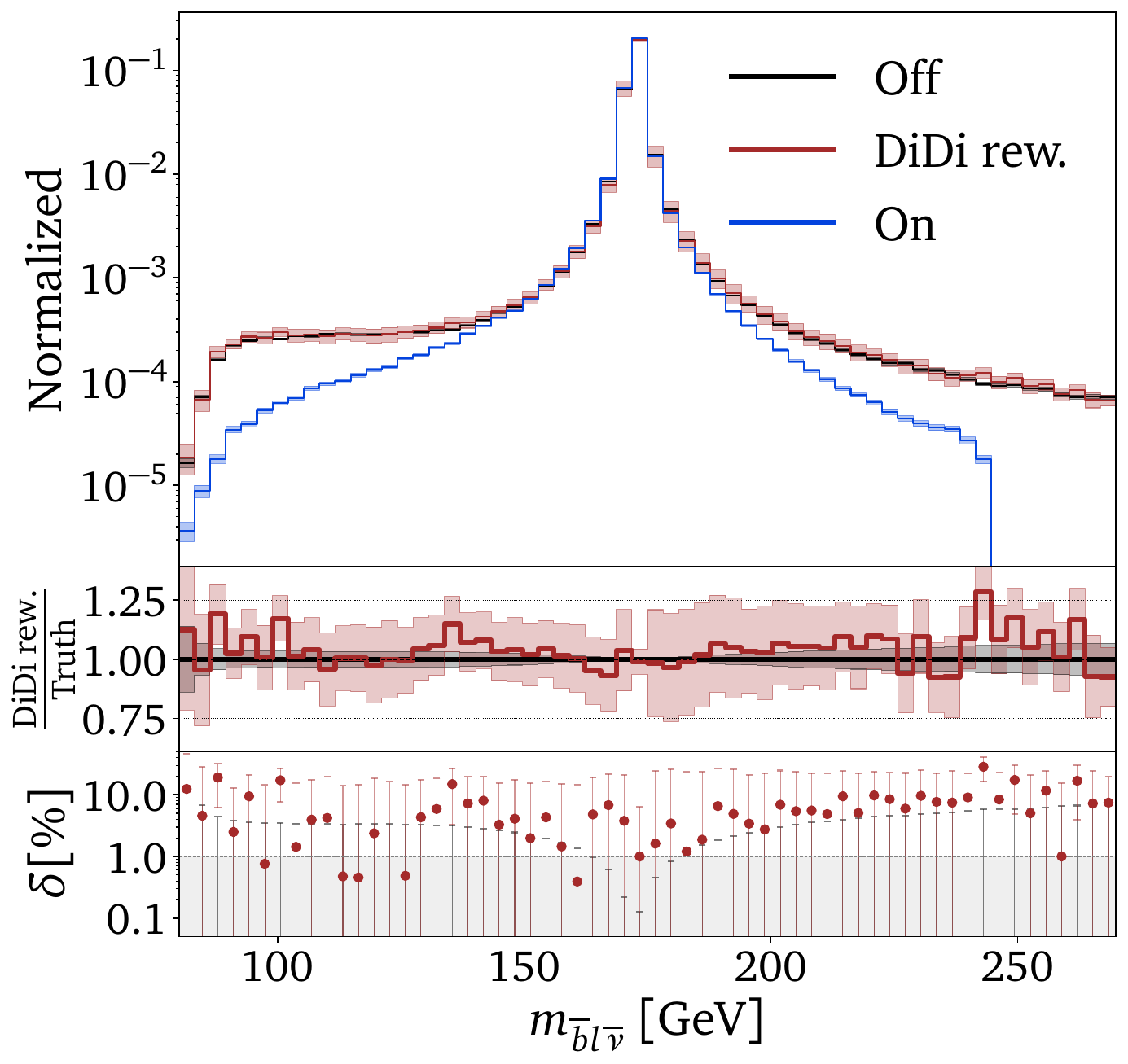}
  \caption{Results from our re-weighted Direct Diffusion off-shell generator,
    compared to the on-shell starting point and the off-shell target distributions.}
    \label{fig:reweighted}
\end{figure}

Even though the shortcomings of our generative networks, visible in
Fig.~\ref{fig:D2F}, arise from noisy network training and do not
reflect systematic shortcomings, they affect the trained generative
network in a systematic, localized manner. As a function of phase space, 
we can correct them using
the event weights from
Eq.\eqref{eq:weights}, because the classifier network is more sensitive and more precise
than the generator network~\cite{Das:2023ktd}.  In Fig.~\ref{fig:reweighted} we show a set of
kinematic distributions for reweighted events, where the uncertainty is given by 
the Bayesian generator. Comparing the agreement
between the reweighted and the target distribution with the unweighted performance from
Fig.~\ref{fig:D2F} we see significant improvements. In the secondary panels we
show the reweighted predictions from the Bayesian generator, allowing 
us to compare the statistical uncertainty from the training data with 
the predictive uncertainty from the generative network. For the entire 
range of $p_{T,\bar{b}}$ the network agrees with the true distribution
within its predictive uncertainties, and almost within the uncertainties
of the training data. For $m_{\bar{l}b}$, DiDi has to cover phase 
space far beyond the on-shell structures, and, again it hardly exceeds
the statistical uncertainties of the training data and provides a 
conservative uncertainty estimate from the Bayesian setup. 

The momentum and the mass of the reconstructed $W^{-}$ match the 
truth perfectly in the bulk, and roughly within the statistics of the 
training data in the tails. The former is important, because it confirms 
the original motivation that the network reproduces the on-shell features
with extremely high precision. The same can be seen for the reconstructed
anti-top mass, where most of the phase space is filled by extrapolating 
from the on-shell distribution and the network learns these new phase
space regions without degrading the precision in the bulk at all.

%----------------------------------------------------------
%----------------------------------------------------------

%%%%%%%%%%%%%%%%%%%%%%%%%%%%%%%%%%%%%%%%%%%%%%%%%%%%%%%%%%%%%%%%%%%%%%%%
\section{Outlook}
\label{sec:outlook}

Sufficiently fast and precise event generation for the HL-LHC is a
major challenge in the coming decade. A perfect
example is the generation of off-shell kinematic configurations for
large backgrounds. Even when they can be computed from
perturbative QFT, with significant investment in CPU
time, they need to be implemented in standard event generators. A
promising strategy to provide them in an amortized manner is
generative surrogate networks. They are
easy to include and ship with event generators and 
enhance the precision of a limited number of events, just like a
fitted function to a statistically limited dataset.

Generating off-shell events is also an exciting problem for modern
machine learning, because it cannot be solved by regressing an
amplitude at a give phase space
point. Instead, it requires a generative surrogate to cover the
off-shell phase space. We proposed a new method, namely to
generate off-shell configurations relative to given on-shell
configurations. Its advantage 
is that the generative network only learns a
controlled deviation from a simple unit transformation. This
simplified task allows us to generate a rich resonance
structure without the usual challenges in network size and precision
training. The approach is enabled by Direct Diffusion (DiDi)
networks, which 
sample from any given
distribution to produce another distribution, in our case transforming
on-shell events into off-shell events.

For top pair production we have shown that already a relatively small
network with limited training effort can reproduce the target
off-shell distributions at the 10\% level. Most importantly,
the DiDi generation does not encounter any issues with learned
features. The challenging on-shell peaks are described at the 5\% level
or better, as seen in Fig.~\ref{fig:D2F}.

Using a classifier reweighting we can improve its precision
to the few per-cent level, even in challenging kinematic
distributions, Fig.~\ref{fig:reweighted}. Now the bulk precision
exceeds 1\%, and the precision in the extrapolation
region is reliably below or at the 10\% level. The same classifier allows us to 
control the performance of the generative networks
and ensures that all local features in phase space are correctly 
learned. Finally, the Bayesian extension of the DiDi 
architecture provides us with a conservative uncertainty estimate,
tracking the effects of statistically limited training data.

Our approach can be extended to higher-orders as long as the corresponding event samples can be provided as training data.
Such an extension is not trivial because with each extra particle in the final state, e.g.~due to an extra emission in the NLO real correction, the dimension of the phase space increases.
We do not foresee any conceptual issues however, as increasing the dimension of the phase space in our setup is straightforward and relatively small event samples were needed at LO, and both diffusion networks as well as Classifiers have been shown to scale well with dimension.
The size of the off-shell effect, or more specifically the ratio of the full and approximate off-shell prediction for a given observable, could in principle change dramatically when changing orders.\footnote{For the example of $t\bar{t}$ production at hand, we compared the size of the off-shell effect at LO vs at NLO, and did not find differences in excess of one order of magnitude.} 
This is not a problem though, as DiDi is able to reproduce full off-shell predictions also in regions where the approximate off-shell prediction is vanishing. 
For realistic predictions we will also need to address issues related to matching to the parton shower, like shower starting scale and colourflow configurations not discussed here. 
For the latter we may have to include single top production in our on-shell data set.

Our fast generative surrogate is ready to be implemented standalone using Les Houches Event (LHE) format for input and output. Generative networks of this kind will also be part of the 
ML-enhanced ultrafast \textsc{Madgraph} event
generator~\cite{Heimel:2022wyj,Heimel:2023ngj}.
%%%%%%%%%%%%%%%%%%%%%%%%%%%%%%%%%%%%%%%%%%%%%%%%%%%%%%%%%%%%%%%%%%%%%%%%
\section*{Acknowledgements}

We thank Nathan Huetsch for many fruitful discussions. AB and TP are supported by the Deutsche Forschungsgemeinschaft (DFG,
German Research Foundation) under grant 396021762 -- TRR~257
\textsl{Particle Physics Phenomenology after the Higgs Discovery}.
AB gratefully acknowledges the continuous support from LPNHE, CNRS/IN2P3, Sorbonne Université and Université de Paris Cité.
Work at the University of M\"unster is supported by the BMBF through
project \textsl{InterKIWWU} and by the DFG through SFB 1225
\textsl{Isoquant}, project-id 273811115, and the Research Training
Group 2149 \textsl{Strong and Weak Interactions - from Hadrons to Dark
  Matter}.  SPS is supported by the BMBF Junior Group \textsl{Generative Precision Networks for Particle Physics} (DLR 01IS22079). The authors acknowledge support by the state of
Baden-W\"urttemberg through bwHPC and the German Research Foundation
(DFG) through grant no INST 39/963-1 FUGG (bwForCluster NEMO).  This
work was supported by the DFG under Germany’s Excellence Strategy EXC
2181/1 - 390900948 \textsl{The Heidelberg STRUCTURES Excellence
  Cluster}.

\clearpage
%%%%%%%%%%%%%%%%%%%%%%%%%%%%%%%%%%%%%%%%%%%%%%%%%%%%%%%%%%%%%%%%%%%%%%%%
\bibliographystyle{tepml}
\bibliography{paper} 

\providecommand{\href}[2]{#2}\begingroup\raggedright\begin{thebibliography}{100}

\bibitem{Butter:2022rso}
A.~Butter, T.~Plehn, S.~Schumann, {\em et al.}, {\it {Machine Learning and LHC
  Event Generation}},  in {\em {2022 Snowmass Summer Study}}.
\newblock 3, 2022.
\newblock
\newblock \href{http://arxiv.org/abs/2203.07460}{{arXiv:2203.07460 [hep-ph]}}.

\bibitem{Cranmer:2019eaq}
K.~Cranmer, J.~Brehmer, and G.~Louppe, {\it {The frontier of simulation-based
  inference}},  \href{http://dx.doi.org/10.1073/pnas.1912789117}{Proc. Nat.
  Acad. Sci. {\bfseries 117} (2020) 48, 30055},
  \href{http://arxiv.org/abs/1911.01429}{{arXiv:1911.01429 [stat.ML]}}.

\bibitem{Campbell:2022qmc}
J.~M. Campbell {\em et al.}, {\it {Event Generators for High-Energy Physics
  Experiments}},  in {\em {2022 Snowmass Summer Study}}.
\newblock 3, 2022.
\newblock
\newblock \href{http://arxiv.org/abs/2203.11110}{{arXiv:2203.11110 [hep-ph]}}.

\bibitem{Heinrich:2017bqp}
G.~Heinrich, A.~Maier, R.~Nisius, J.~Schlenk, M.~Schulze, L.~Scyboz, and
  J.~Winter, {\it {NLO and off-shell effects in top quark mass
  determinations}},  \href{http://dx.doi.org/10.1007/JHEP07(2018)129}{JHEP
  {\bfseries 07} (2018)  129},
  \href{http://arxiv.org/abs/1709.08615}{{arXiv:1709.08615 [hep-ph]}}.

\bibitem{FerrarioRavasio:2018whr}
S.~Ferrario~Ravasio, T.~Je\v{z}o, P.~Nason, and C.~Oleari, {\it {A theoretical
  study of top-mass measurements at the LHC using NLO+PS generators of
  increasing accuracy}},
  \href{http://dx.doi.org/10.1140/epjc/s10052-019-7336-9}{Eur. Phys. J. C
  {\bfseries 78} (2018) 6, 458},
  \href{http://arxiv.org/abs/1906.09166}{{arXiv:1906.09166 [hep-ph]}}.
  [Addendum: Eur.Phys.J.C 79, 859 (2019)].

\bibitem{Plehn:2022ftl}
T.~Plehn, A.~Butter, B.~Dillon, and C.~Krause, {\it {Modern Machine Learning
  for LHC Physicists}},
  \href{http://arxiv.org/abs/2211.01421}{{arXiv:2211.01421 [hep-ph]}}.

\bibitem{Bishara:2019iwh}
F.~Bishara and M.~Montull, {\it {(Machine) Learning Amplitudes for Faster Event
  Generation}},   \href{http://arxiv.org/abs/1912.11055}{{arXiv:1912.11055
  [hep-ph]}}.

\bibitem{Badger:2020uow}
S.~Badger and J.~Bullock, {\it {Using neural networks for efficient evaluation
  of high multiplicity scattering amplitudes}},
  \href{http://dx.doi.org/10.1007/JHEP06(2020)114}{JHEP {\bfseries 06} (2020)
  114},  \href{http://arxiv.org/abs/2002.07516}{{arXiv:2002.07516 [hep-ph]}}.

\bibitem{Aylett-Bullock:2021hmo}
J.~Aylett-Bullock, S.~Badger, and R.~Moodie, {\it {Optimising simulations for
  diphoton production at hadron colliders using amplitude neural networks}},
  \href{http://dx.doi.org/10.1007/JHEP08(2021)066}{JHEP {\bfseries 08} (2021)
  066},  \href{http://arxiv.org/abs/2106.09474}{{arXiv:2106.09474 [hep-ph]}}.

\bibitem{Maitre:2021uaa}
D.~Ma\^\i{}tre and H.~Truong, {\it {A factorisation-aware Matrix element
  emulator}},  \href{http://dx.doi.org/10.1007/JHEP11(2021)066}{JHEP {\bfseries
  11} (2021)  066},  \href{http://arxiv.org/abs/2107.06625}{{arXiv:2107.06625
  [hep-ph]}}.

\bibitem{Winterhalder:2021ngy}
R.~Winterhalder, V.~Magerya, E.~Villa, S.~P. Jones, M.~Kerner, A.~Butter,
  G.~Heinrich, and T.~Plehn, {\it {Targeting multi-loop integrals with neural
  networks}},  \href{http://dx.doi.org/10.21468/SciPostPhys.12.4.129}{SciPost
  Phys. {\bfseries 12} (2022) 4, 129},
  \href{http://arxiv.org/abs/2112.09145}{{arXiv:2112.09145 [hep-ph]}}.

\bibitem{Badger:2022hwf}
S.~Badger, A.~Butter, M.~Luchmann, S.~Pitz, and T.~Plehn, {\it {Loop Amplitudes
  from Precision Networks}},
  \href{http://arxiv.org/abs/2206.14831}{{arXiv:2206.14831 [hep-ph]}}.

\bibitem{Maitre:2023dqz}
D.~Ma\^\i{}tre and H.~Truong, {\it {One-loop matrix element emulation with
  factorisation awareness}},
  \href{http://arxiv.org/abs/2302.04005}{{arXiv:2302.04005 [hep-ph]}}.

\bibitem{Hagiwara:2013oka}
K.~Hagiwara, J.~Kanzaki, Q.~Li, N.~Okamura, and T.~Stelzer, {\it {Fast
  computation of MadGraph amplitudes on graphics processing unit (GPU)}},
  \href{http://dx.doi.org/10.1140/epjc/s10052-013-2608-2}{Eur. Phys. J. C
  {\bfseries 73} (2013)  2608},
  \href{http://arxiv.org/abs/1305.0708}{{arXiv:1305.0708 [physics.comp-ph]}}.

\bibitem{Bothmann:2021nch}
E.~Bothmann, W.~Giele, S.~Hoeche, J.~Isaacson, and M.~Knobbe, {\it {Many-gluon
  tree amplitudes on modern GPUs: A case study for novel event generators}},
  \href{http://arxiv.org/abs/2106.06507}{{arXiv:2106.06507 [hep-ph]}}.

\bibitem{Carrazza:2021gpx}
S.~Carrazza, J.~Cruz-Martinez, M.~Rossi, and M.~Zaro, {\it {MadFlow: automating
  Monte Carlo simulation on GPU for particle physics processes}},
  \href{http://dx.doi.org/10.1140/epjc/s10052-021-09443-8}{Eur. Phys. J. C
  {\bfseries 81} (2021) 7, 656},
  \href{http://arxiv.org/abs/2106.10279}{{arXiv:2106.10279 [physics.comp-ph]}}.

\bibitem{Valassi:2021ljk}
A.~Valassi, S.~Roiser, O.~Mattelaer, and S.~Hageboeck, {\it {Design and
  engineering of a simplified workflow execution for the MG5aMC event generator
  on GPUs and vector CPUs}},
  \href{http://dx.doi.org/10.1051/epjconf/202125103045}{EPJ Web Conf.
  {\bfseries 251} (2021)  03045},
  \href{http://arxiv.org/abs/2106.12631}{{arXiv:2106.12631 [physics.comp-ph]}}.

\bibitem{Danziger:2021eeg}
K.~Danziger, T.~Jan\ss{}en, S.~Schumann, and F.~Siegert, {\it {Accelerating
  Monte Carlo event generation -- rejection sampling using neural network
  event-weight estimates}},
  \href{http://arxiv.org/abs/2109.11964}{{arXiv:2109.11964 [hep-ph]}}.

\bibitem{Chen:2020nfb}
I.-K. Chen, M.~D. Klimek, and M.~Perelstein, {\it {Improved Neural Network
  Monte Carlo Simulation}},
  \href{http://dx.doi.org/10.21468/SciPostPhys.10.1.023}{SciPost Phys.
  {\bfseries 10} (2021)  023},
  \href{http://arxiv.org/abs/2009.07819}{{arXiv:2009.07819 [hep-ph]}}.

\bibitem{Gao:2020vdv}
C.~Gao, J.~Isaacson, and C.~Krause, {\it {i-flow: High-dimensional Integration
  and Sampling with Normalizing Flows}},
  \href{http://dx.doi.org/10.1088/2632-2153/abab62}{Mach. Learn. Sci. Tech.
  {\bfseries 1} (2020) 4, 045023},
  \href{http://arxiv.org/abs/2001.05486}{{arXiv:2001.05486 [physics.comp-ph]}}.

\bibitem{Bothmann:2020ywa}
E.~Bothmann, T.~Jan{\ss}en, M.~Knobbe, T.~Schmale, and S.~Schumann, {\it
  {Exploring phase space with Neural Importance Sampling}},
  \href{http://dx.doi.org/10.21468/SciPostPhys.8.4.069}{SciPost Phys.
  {\bfseries 8} (2020) 4, 069},
  \href{http://arxiv.org/abs/2001.05478}{{arXiv:2001.05478 [hep-ph]}}.

\bibitem{Gao:2020zvv}
C.~Gao, S.~Höche, J.~Isaacson, C.~Krause, and H.~Schulz, {\it {Event
  Generation with Normalizing Flows}},
  \href{http://dx.doi.org/10.1103/PhysRevD.101.076002}{Phys. Rev. D {\bfseries
  101} (2020) 7, 076002},
  \href{http://arxiv.org/abs/2001.10028}{{arXiv:2001.10028 [hep-ph]}}.

\bibitem{Heimel:2022wyj}
T.~Heimel, R.~Winterhalder, A.~Butter, J.~Isaacson, C.~Krause, F.~Maltoni,
  O.~Mattelaer, and T.~Plehn, {\it {MadNIS -- Neural Multi-Channel Importance
  Sampling}},   \href{http://arxiv.org/abs/2212.06172}{{arXiv:2212.06172
  [hep-ph]}}.

\bibitem{Heimel:2023ngj}
T.~Heimel, N.~Huetsch, F.~Maltoni, O.~Mattelaer, T.~Plehn, and R.~Winterhalder,
  {\it {The MadNIS Reloaded}},
  \href{http://arxiv.org/abs/2311.01548}{{arXiv:2311.01548 [hep-ph]}}.

\bibitem{Janssen:2023ahv}
T.~Jan\ss{}en, D.~Ma\^\i{}tre, S.~Schumann, F.~Siegert, and H.~Truong, {\it
  {Unweighting multijet event generation using factorisation-aware neural
  networks}},  \href{http://dx.doi.org/10.21468/SciPostPhys.15.3.107}{SciPost
  Phys. {\bfseries 15} (2023)  107},
  \href{http://arxiv.org/abs/2301.13562}{{arXiv:2301.13562 [hep-ph]}}.

\bibitem{Bothmann:2023siu}
E.~Bothmann, T.~Childers, W.~Giele, F.~Herren, S.~Hoeche, J.~Isaacson,
  M.~Knobbe, and R.~Wang, {\it {Efficient phase-space generation for hadron
  collider event simulation}},
  \href{http://dx.doi.org/10.21468/SciPostPhys.15.4.169}{SciPost Phys.
  {\bfseries 15} (2023) 4, 169},
  \href{http://arxiv.org/abs/2302.10449}{{arXiv:2302.10449 [hep-ph]}}.

\bibitem{dutch}
S.~Otten, S.~Caron, W.~de~Swart, M.~van Beekveld, L.~Hendriks, C.~van Leeuwen,
  D.~Podareanu, R.~Ruiz~de Austri, and R.~Verheyen, {\it {Event Generation and
  Statistical Sampling for Physics with Deep Generative Models and a Density
  Information Buffer}},
  \href{http://dx.doi.org/10.1038/s41467-021-22616-z}{Nature Commun. {\bfseries
  12} (2021) 1, 2985},
  \href{http://arxiv.org/abs/1901.00875}{{arXiv:1901.00875 [hep-ph]}}.

\bibitem{gan_datasets}
B.~Hashemi, N.~Amin, K.~Datta, D.~Olivito, and M.~Pierini, {\it {LHC
  analysis-specific datasets with Generative Adversarial Networks}},
  \href{http://arxiv.org/abs/1901.05282}{{arXiv:1901.05282 [hep-ex]}}.

\bibitem{DijetGAN2}
R.~Di~Sipio, M.~Faucci~Giannelli, S.~Ketabchi~Haghighat, and S.~Palazzo, {\it
  {DijetGAN: A Generative-Adversarial Network Approach for the Simulation of
  QCD Dijet Events at the LHC}},
  \href{http://dx.doi.org/10.1007/JHEP08(2019)110}{JHEP {\bfseries 08} (2020)
  110},
\href{http://arxiv.org/abs/1903.02433}{{arXiv:1903.02433 [hep-ex]}}.
%%CITATION = ARXIV:1903.02433;%%.

\bibitem{Butter:2019cae}
A.~Butter, T.~Plehn, and R.~Winterhalder, {\it {How to GAN LHC Events}},
  \href{http://dx.doi.org/10.21468/SciPostPhys.7.6.075}{SciPost Phys.
  {\bfseries 7} (2019) 6, 075},
  \href{http://arxiv.org/abs/1907.03764}{{arXiv:1907.03764 [hep-ph]}}.

\bibitem{Alanazi:2020klf}
Y.~Alanazi, N.~Sato, T.~Liu, W.~Melnitchouk, M.~P. Kuchera, E.~Pritchard,
  M.~Robertson, R.~Strauss, L.~Velasco, and Y.~Li, {\it {Simulation of
  electron-proton scattering events by a Feature-Augmented and Transformed
  Generative Adversarial Network (FAT-GAN)}},
  \href{http://arxiv.org/abs/2001.11103}{{arXiv:2001.11103 [hep-ph]}}.

\bibitem{Butter:2021csz}
A.~Butter, T.~Heimel, S.~Hummerich, T.~Krebs, T.~Plehn, A.~Rousselot, and
  S.~Vent, {\it {Generative Networks for Precision Enthusiasts}},
  \href{http://arxiv.org/abs/2110.13632}{{arXiv:2110.13632 [hep-ph]}}.

\bibitem{Butter:2023fov}
A.~Butter, N.~Huetsch, S.~Palacios~Schweitzer, T.~Plehn, P.~Sorrenson, and
  J.~Spinner, {\it {Jet Diffusion versus JetGPT -- Modern Networks for the
  LHC}},   \href{http://arxiv.org/abs/2305.10475}{{arXiv:2305.10475 [hep-ph]}}.

\bibitem{locationGAN}
L.~de~Oliveira, M.~Paganini, and B.~Nachman, {\it {Learning Particle Physics by
  Example: Location-Aware Generative Adversarial Networks for Physics
  Synthesis}},  \href{http://dx.doi.org/10.1007/s41781-017-0004-6}{Comput.
  Softw. Big Sci. {\bfseries 1} (2017) 1, 4},
\href{http://arxiv.org/abs/1701.05927}{{arXiv:1701.05927 [stat.ML]}}.
%%CITATION = ARXIV:1701.05927;%%.

\bibitem{Andreassen:2018apy}
A.~Andreassen, I.~Feige, C.~Frye, and M.~D. Schwartz, {\it {JUNIPR: a Framework
  for Unsupervised Machine Learning in Particle Physics}},
  \href{http://dx.doi.org/10.1140/epjc/s10052-019-6607-9}{Eur. Phys. J.
  {\bfseries C79} (2019) 2, 102},
\href{http://arxiv.org/abs/1804.09720}{{arXiv:1804.09720 [hep-ph]}}.
%%CITATION = ARXIV:1804.09720;%%.

\bibitem{Bothmann:2018trh}
E.~Bothmann and L.~Debbio, {\it {Reweighting a parton shower using a neural
  network: the final-state case}},
  \href{http://dx.doi.org/10.1007/JHEP01(2019)033}{JHEP {\bfseries 01} (2019)
  033},  \href{http://arxiv.org/abs/1808.07802}{{arXiv:1808.07802 [hep-ph]}}.

\bibitem{Dohi:2020eda}
K.~Dohi, {\it {Variational Autoencoders for Jet Simulation}},
  \href{http://arxiv.org/abs/2009.04842}{{arXiv:2009.04842 [hep-ph]}}.

\bibitem{Buhmann:2023pmh}
E.~Buhmann, G.~Kasieczka, and J.~Thaler, {\it {EPiC-GAN: Equivariant Point
  Cloud Generation for Particle Jets}},
  \href{http://dx.doi.org/10.21468/SciPostPhys.15.4.130}{SciPost Phys.
  {\bfseries 15} (2023)  130},
  \href{http://arxiv.org/abs/2301.08128}{{arXiv:2301.08128 [hep-ph]}}.

\bibitem{Leigh:2023toe}
M.~Leigh, D.~Sengupta, G.~Qu\'etant, J.~A. Raine, K.~Zoch, and T.~Golling, {\it
  {PC-JeDi: Diffusion for Particle Cloud Generation in High Energy Physics}},
  \href{http://arxiv.org/abs/2303.05376}{{arXiv:2303.05376 [hep-ph]}}.

\bibitem{Mikuni:2023dvk}
V.~Mikuni, B.~Nachman, and M.~Pettee, {\it {Fast point cloud generation with
  diffusion models in high energy physics}},
  \href{http://dx.doi.org/10.1103/PhysRevD.108.036025}{Phys. Rev. D {\bfseries
  108} (2023) 3, 036025},
  \href{http://arxiv.org/abs/2304.01266}{{arXiv:2304.01266 [hep-ph]}}.

\bibitem{Buhmann:2023zgc}
E.~Buhmann, C.~Ewen, D.~A. Faroughy, T.~Golling, G.~Kasieczka, M.~Leigh,
  G.~Qu\'etant, J.~A. Raine, D.~Sengupta, and D.~Shih, {\it {EPiC-ly Fast
  Particle Cloud Generation with Flow-Matching and Diffusion}},
  \href{http://arxiv.org/abs/2310.00049}{{arXiv:2310.00049 [hep-ph]}}.

\bibitem{Paganini:2017hrr}
M.~Paganini, L.~de~Oliveira, and B.~Nachman, {\it {Accelerating Science with
  Generative Adversarial Networks: An Application to 3D Particle Showers in
  Multilayer Calorimeters}},
  \href{http://dx.doi.org/10.1103/PhysRevLett.120.042003}{Phys. Rev. Lett.
  {\bfseries 120} (2018) 4, 042003},
  \href{http://arxiv.org/abs/1705.02355}{{arXiv:1705.02355 [hep-ex]}}.

\bibitem{deOliveira:2017rwa}
L.~de~Oliveira, M.~Paganini, and B.~Nachman, {\it {Controlling Physical
  Attributes in GAN-Accelerated Simulation of Electromagnetic Calorimeters}},
  \href{http://dx.doi.org/10.1088/1742-6596/1085/4/042017}{J. Phys. Conf. Ser.
  {\bfseries 1085} (2018) 4, 042017},
  \href{http://arxiv.org/abs/1711.08813}{{arXiv:1711.08813 [hep-ex]}}.

\bibitem{Paganini:2017dwg}
M.~Paganini, L.~de~Oliveira, and B.~Nachman, {\it {CaloGAN : Simulating 3D high
  energy particle showers in multilayer electromagnetic calorimeters with
  generative adversarial networks}},
  \href{http://dx.doi.org/10.1103/PhysRevD.97.014021}{Phys. Rev. D {\bfseries
  97} (2018) 1, 014021},
  \href{http://arxiv.org/abs/1712.10321}{{arXiv:1712.10321 [hep-ex]}}.

\bibitem{Erdmann:2018kuh}
M.~Erdmann, L.~Geiger, J.~Glombitza, and D.~Schmidt, {\it {Generating and
  refining particle detector simulations using the Wasserstein distance in
  adversarial networks}},
  \href{http://dx.doi.org/10.1007/s41781-018-0008-x}{Comput. Softw. Big Sci.
  {\bfseries 2} (2018) 1, 4},
  \href{http://arxiv.org/abs/1802.03325}{{arXiv:1802.03325 [astro-ph.IM]}}.

\bibitem{Erdmann:2018jxd}
M.~Erdmann, J.~Glombitza, and T.~Quast, {\it {Precise simulation of
  electromagnetic calorimeter showers using a Wasserstein Generative
  Adversarial Network}},
  \href{http://dx.doi.org/10.1007/s41781-018-0019-7}{Comput. Softw. Big Sci.
  {\bfseries 3} (2019) 1, 4},
  \href{http://arxiv.org/abs/1807.01954}{{arXiv:1807.01954 [physics.ins-det]}}.

\bibitem{Belayneh:2019vyx}
D.~Belayneh {\em et al.}, {\it {Calorimetry with Deep Learning: Particle
  Simulation and Reconstruction for Collider Physics}},
  \href{http://dx.doi.org/10.1140/epjc/s10052-020-8251-9}{Eur. Phys. J. C
  {\bfseries 80} (2020) 7, 688},
  \href{http://arxiv.org/abs/1912.06794}{{arXiv:1912.06794 [physics.ins-det]}}.

\bibitem{Buhmann:2020pmy}
E.~Buhmann, S.~Diefenbacher, E.~Eren, F.~Gaede, G.~Kasieczka, A.~Korol, and
  K.~Kr\"uger, {\it {Getting High: High Fidelity Simulation of High Granularity
  Calorimeters with High Speed}},
  \href{http://dx.doi.org/10.1007/s41781-021-00056-0}{Comput. Softw. Big Sci.
  {\bfseries 5} (2021) 1, 13},
  \href{http://arxiv.org/abs/2005.05334}{{arXiv:2005.05334 [physics.ins-det]}}.

\bibitem{Buhmann:2021lxj}
E.~Buhmann, S.~Diefenbacher, E.~Eren, F.~Gaede, G.~Kasieczka, A.~Korol, and
  K.~Kr\"uger, {\it {Decoding Photons: Physics in the Latent Space of a BIB-AE
  Generative Network}},
  \href{http://arxiv.org/abs/2102.12491}{{arXiv:2102.12491 [physics.ins-det]}}.

\bibitem{Krause:2021ilc}
C.~Krause and D.~Shih, {\it {CaloFlow: Fast and Accurate Generation of
  Calorimeter Showers with Normalizing Flows}},
  \href{http://arxiv.org/abs/2106.05285}{{arXiv:2106.05285 [physics.ins-det]}}.

\bibitem{ATLAS:2021pzo}
{ATLAS Collaboration}, {\it {AtlFast3: the next generation of fast simulation
  in ATLAS}},  \href{http://dx.doi.org/10.1007/s41781-021-00079-7}{Comput.
  Softw. Big Sci. {\bfseries 6} (2022)  7},
  \href{http://arxiv.org/abs/2109.02551}{{arXiv:2109.02551 [hep-ex]}}.

\bibitem{Krause:2021wez}
C.~Krause and D.~Shih, {\it {CaloFlow II: Even Faster and Still Accurate
  Generation of Calorimeter Showers with Normalizing Flows}},
  \href{http://arxiv.org/abs/2110.11377}{{arXiv:2110.11377 [physics.ins-det]}}.

\bibitem{Buhmann:2021caf}
E.~Buhmann, S.~Diefenbacher, D.~Hundhausen, G.~Kasieczka, W.~Korcari, E.~Eren,
  F.~Gaede, K.~Kr\"uger, P.~McKeown, and L.~Rustige, {\it {Hadrons, better,
  faster, stronger}},  \href{http://dx.doi.org/10.1088/2632-2153/ac7848}{Mach.
  Learn. Sci. Tech. {\bfseries 3} (2022) 2, 025014},
  \href{http://arxiv.org/abs/2112.09709}{{arXiv:2112.09709 [physics.ins-det]}}.

\bibitem{Chen:2021gdz}
C.~Chen, O.~Cerri, T.~Q. Nguyen, J.~R. Vlimant, and M.~Pierini, {\it
  {Analysis-Specific Fast Simulation at the LHC with Deep Learning}},
  \href{http://dx.doi.org/10.1007/s41781-021-00060-4}{Comput. Softw. Big Sci.
  {\bfseries 5} (2021) 1, 15}.

\bibitem{Mikuni:2022xry}
V.~Mikuni and B.~Nachman, {\it {Score-based generative models for calorimeter
  shower simulation}},
  \href{http://dx.doi.org/10.1103/PhysRevD.106.092009}{Phys. Rev. D {\bfseries
  106} (2022) 9, 092009},
  \href{http://arxiv.org/abs/2206.11898}{{arXiv:2206.11898 [hep-ph]}}.

\bibitem{ATLAS:2022jhk}
{ATLAS Collaboration}, {\it {Deep generative models for fast photon shower
  simulation in ATLAS}},
  \href{http://arxiv.org/abs/2210.06204}{{arXiv:2210.06204 [hep-ex]}}.

\bibitem{Krause:2022jna}
C.~Krause, I.~Pang, and D.~Shih, {\it {CaloFlow for CaloChallenge Dataset 1}},
   \href{http://arxiv.org/abs/2210.14245}{{arXiv:2210.14245
  [physics.ins-det]}}.

\bibitem{Cresswell:2022tof}
J.~C. Cresswell, B.~L. Ross, G.~Loaiza-Ganem, H.~Reyes-Gonzalez, M.~Letizia,
  and A.~L. Caterini, {\it {CaloMan: Fast generation of calorimeter showers
  with density estimation on learned manifolds}},  in {\em {36th Conference on
  Neural Information Processing Systems}}.
\newblock 11, 2022.
\newblock
\newblock \href{http://arxiv.org/abs/2211.15380}{{arXiv:2211.15380 [hep-ph]}}.

\bibitem{Diefenbacher:2023vsw}
S.~Diefenbacher, E.~Eren, F.~Gaede, G.~Kasieczka, C.~Krause, I.~Shekhzadeh, and
  D.~Shih, {\it {L2LFlows: Generating High-Fidelity 3D Calorimeter Images}},
  \href{http://arxiv.org/abs/2302.11594}{{arXiv:2302.11594 [physics.ins-det]}}.

\bibitem{Hashemi:2023ruu}
H.~Hashemi, N.~Hartmann, S.~Sharifzadeh, J.~Kahn, and T.~Kuhr, {\it
  {Ultra-High-Resolution Detector Simulation with Intra-Event Aware GAN and
  Self-Supervised Relational Reasoning}},
  \href{http://arxiv.org/abs/2303.08046}{{arXiv:2303.08046 [physics.ins-det]}}.

\bibitem{Xu:2023xdc}
A.~Xu, S.~Han, X.~Ju, and H.~Wang, {\it {Generative Machine Learning for
  Detector Response Modeling with a Conditional Normalizing Flow}},
  \href{http://arxiv.org/abs/2303.10148}{{arXiv:2303.10148 [hep-ex]}}.

\bibitem{Diefenbacher:2023prl}
S.~Diefenbacher, E.~Eren, F.~Gaede, G.~Kasieczka, A.~Korol, K.~Kr\"uger,
  P.~McKeown, and L.~Rustige, {\it {New angles on fast calorimeter shower
  simulation}},  \href{http://dx.doi.org/10.1088/2632-2153/acefa9}{Mach. Learn.
  Sci. Tech. {\bfseries 4} (2023) 3, 035044},
  \href{http://arxiv.org/abs/2303.18150}{{arXiv:2303.18150 [physics.ins-det]}}.

\bibitem{Buhmann:2023bwk}
E.~Buhmann, S.~Diefenbacher, E.~Eren, F.~Gaede, G.~Kasieczka, A.~Korol,
  W.~Korcari, K.~Kr\"uger, and P.~McKeown, {\it {CaloClouds: Fast
  Geometry-Independent Highly-Granular Calorimeter Simulation}},
  \href{http://arxiv.org/abs/2305.04847}{{arXiv:2305.04847 [physics.ins-det]}}.

\bibitem{Buckley:2023rez}
M.~R. Buckley, C.~Krause, I.~Pang, and D.~Shih, {\it {Inductive CaloFlow}},
  \href{http://arxiv.org/abs/2305.11934}{{arXiv:2305.11934 [physics.ins-det]}}.

\bibitem{Diefenbacher:2023flw}
S.~Diefenbacher, V.~Mikuni, and B.~Nachman, {\it {Refining Fast Calorimeter
  Simulations with a Schr\"odinger Bridge}},
  \href{http://arxiv.org/abs/2308.12339}{{arXiv:2308.12339 [physics.ins-det]}}.

\bibitem{Golling:2023mqx}
T.~Golling, S.~Klein, R.~Mastandrea, B.~Nachman, and J.~A. Raine, {\it {Flows
  for Flows: Morphing one Dataset into another with Maximum Likelihood
  Estimation}},   \href{http://arxiv.org/abs/2309.06472}{{arXiv:2309.06472
  [hep-ph]}}.

\bibitem{Diefenbacher:2023wec}
S.~Diefenbacher, G.-H. Liu, V.~Mikuni, B.~Nachman, and W.~Nie, {\it {Improving
  Generative Model-based Unfolding with Schr\"odinger Bridges}},
  \href{http://arxiv.org/abs/2308.12351}{{arXiv:2308.12351 [hep-ph]}}.

\bibitem{Winterhalder:2021ave}
R.~Winterhalder, M.~Bellagente, and B.~Nachman, {\it {Latent Space Refinement
  for Deep Generative Models}},
  \href{http://arxiv.org/abs/2106.00792}{{arXiv:2106.00792 [stat.ML]}}.

\bibitem{Nachman:2023clf}
B.~Nachman and R.~Winterhalder, {\it {Elsa: enhanced latent spaces for improved
  collider simulations}},
  \href{http://dx.doi.org/10.1140/epjc/s10052-023-11989-8}{Eur. Phys. J. C
  {\bfseries 83} (2023) 9, 843},
  \href{http://arxiv.org/abs/2305.07696}{{arXiv:2305.07696 [hep-ph]}}.

\bibitem{Leigh:2023zle}
M.~Leigh, D.~Sengupta, J.~A. Raine, G.~Qu\'etant, and T.~Golling, {\it
  {PC-Droid: Faster diffusion and improved quality for particle cloud
  generation}},   \href{http://arxiv.org/abs/2307.06836}{{arXiv:2307.06836
  [hep-ex]}}.

\bibitem{Das:2023ktd}
R.~Das, L.~Favaro, T.~Heimel, C.~Krause, T.~Plehn, and D.~Shih, {\it {How to
  Understand Limitations of Generative Networks}},
  \href{http://arxiv.org/abs/2305.16774}{{arXiv:2305.16774 [hep-ph]}}.

\bibitem{Butter:2020qhk}
A.~Butter, S.~Diefenbacher, G.~Kasieczka, B.~Nachman, and T.~Plehn, {\it
  {GANplifying event samples}},
  \href{http://dx.doi.org/10.21468/SciPostPhys.10.6.139}{SciPost Phys.
  {\bfseries 10} (2021) 6, 139},
  \href{http://arxiv.org/abs/2008.06545}{{arXiv:2008.06545 [hep-ph]}}.

\bibitem{Bieringer:2022cbs}
S.~Bieringer, A.~Butter, S.~Diefenbacher, E.~Eren, F.~Gaede, D.~Hundhausen,
  G.~Kasieczka, B.~Nachman, T.~Plehn, and M.~Trabs, {\it {Calomplification
  \textemdash{} the power of generative calorimeter models}},
  \href{http://dx.doi.org/10.1088/1748-0221/17/09/P09028}{JINST {\bfseries 17}
  (2022) 09, P09028},  \href{http://arxiv.org/abs/2202.07352}{{arXiv:2202.07352
  [hep-ph]}}.

\bibitem{Butter:2019eyo}
A.~Butter, T.~Plehn, and R.~Winterhalder, {\it {How to GAN Event Subtraction}},
   \href{http://dx.doi.org/10.21468/SciPostPhysCore.3.2.009}{SciPost Phys. Core
  {\bfseries 3} (2020)  009},
  \href{http://arxiv.org/abs/1912.08824}{{arXiv:1912.08824 [hep-ph]}}.

\bibitem{Verheyen:2020bjw}
B.~Stienen and R.~Verheyen, {\it {Phase Space Sampling and Inference from
  Weighted Events with Autoregressive Flows}},
  \href{http://dx.doi.org/10.21468/SciPostPhys.10.2.038}{SciPost Phys.
  {\bfseries 10} (2021)  038},
  \href{http://arxiv.org/abs/2011.13445}{{arXiv:2011.13445 [hep-ph]}}.

\bibitem{Backes:2020vka}
M.~Backes, A.~Butter, T.~Plehn, and R.~Winterhalder, {\it {How to GAN Event
  Unweighting}},
  \href{http://dx.doi.org/10.21468/SciPostPhys.10.4.089}{SciPost Phys.
  {\bfseries 10} (2021) 4, 089},
  \href{http://arxiv.org/abs/2012.07873}{{arXiv:2012.07873 [hep-ph]}}.

\bibitem{DiBello:2020bas}
F.~A. Di~Bello, S.~Ganguly, E.~Gross, M.~Kado, M.~Pitt, L.~Santi, and
  J.~Shlomi, {\it {Towards a Computer Vision Particle Flow}},
  \href{http://dx.doi.org/10.1140/epjc/s10052-021-08897-0}{Eur. Phys. J. C
  {\bfseries 81} (2021) 2, 107},
  \href{http://arxiv.org/abs/2003.08863}{{arXiv:2003.08863 [physics.data-an]}}.

\bibitem{Baldi:2020hjm}
P.~Baldi, L.~Blecher, A.~Butter, J.~Collado, J.~N. Howard, F.~Keilbach,
  T.~Plehn, G.~Kasieczka, and D.~Whiteson, {\it {How to GAN Higher Jet
  Resolution}},   \href{http://arxiv.org/abs/2012.11944}{{arXiv:2012.11944
  [hep-ph]}}.

\bibitem{Datta:2018mwd}
K.~Datta, D.~Kar, and D.~Roy, {\it {Unfolding with Generative Adversarial
  Networks}},
\href{http://arxiv.org/abs/1806.00433}{{arXiv:1806.00433 [physics.data-an]}}.
%%CITATION = ARXIV:1806.00433;%%.

\bibitem{Bellagente:2019uyp}
M.~Bellagente, A.~Butter, G.~Kasieczka, T.~Plehn, and R.~Winterhalder, {\it
  {How to GAN away Detector Effects}},
\href{http://arxiv.org/abs/1912.00477}{{arXiv:1912.00477 [hep-ph]}}.
%%CITATION = ARXIV:1912.00477;%%.

\bibitem{Andreassen:2019cjw}
A.~Andreassen, P.~T. Komiske, E.~M. Metodiev, B.~Nachman, and J.~Thaler, {\it
  {OmniFold: A Method to Simultaneously Unfold All Observables}},
  \href{http://dx.doi.org/10.1103/PhysRevLett.124.182001}{Phys. Rev. Lett.
  {\bfseries 124} (2020) 18, 182001},
  \href{http://arxiv.org/abs/1911.09107}{{arXiv:1911.09107 [hep-ph]}}.

\bibitem{Bellagente:2020piv}
M.~Bellagente, A.~Butter, G.~Kasieczka, T.~Plehn, A.~Rousselot,
  R.~Winterhalder, L.~Ardizzone, and U.~K\"othe, {\it {Invertible Networks or
  Partons to Detector and Back Again}},
  \href{http://dx.doi.org/10.21468/SciPostPhys.9.5.074}{SciPost Phys.
  {\bfseries 9} (2020)  074},
  \href{http://arxiv.org/abs/2006.06685}{{arXiv:2006.06685 [hep-ph]}}.

\bibitem{Backes:2022vmn}
M.~Backes, A.~Butter, M.~Dunford, and B.~Malaescu, {\it {An unfolding method
  based on conditional Invertible Neural Networks (cINN) using iterative
  training}},   \href{http://arxiv.org/abs/2212.08674}{{arXiv:2212.08674
  [hep-ph]}}.

\bibitem{Leigh:2022lpn}
M.~Leigh, J.~A. Raine, K.~Zoch, and T.~Golling, {\it {\ensuremath{\nu}-Flows:
  Conditional Neutrino Regression}},
  \href{http://dx.doi.org/10.21468/SciPostPhys.14.6.159}{SciPost Phys.
  {\bfseries 14} (2023)  159},
  \href{http://arxiv.org/abs/2207.00664}{{arXiv:2207.00664 [hep-ph]}}.

\bibitem{Raine:2023fko}
J.~A. Raine, M.~Leigh, K.~Zoch, and T.~Golling, {\it {$\nu^2$-Flows: Fast and
  improved neutrino reconstruction in multi-neutrino final states with
  conditional normalizing flows}},
  \href{http://arxiv.org/abs/2307.02405}{{arXiv:2307.02405 [hep-ph]}}.

\bibitem{Shmakov:2023kjj}
A.~Shmakov, K.~Greif, M.~Fenton, A.~Ghosh, P.~Baldi, and D.~Whiteson, {\it
  {End-To-End Latent Variational Diffusion Models for Inverse Problems in High
  Energy Physics}},   \href{http://arxiv.org/abs/2305.10399}{{arXiv:2305.10399
  [hep-ex]}}.

\bibitem{Ackerschott:2023nax}
J.~Ackerschott, R.~K. Barman, D.~Gon\c{c}alves, T.~Heimel, and T.~Plehn, {\it
  {Returning CP-Observables to The Frames They Belong}},
  \href{http://arxiv.org/abs/2308.00027}{{arXiv:2308.00027 [hep-ph]}}.

\bibitem{Bieringer:2020tnw}
S.~Bieringer, A.~Butter, T.~Heimel, S.~H\"oche, U.~K\"othe, T.~Plehn, and S.~T.
  Radev, {\it {Measuring QCD Splittings with Invertible Networks}},
  \href{http://dx.doi.org/10.21468/SciPostPhys.10.6.126}{SciPost Phys.
  {\bfseries 10} (2021) 6, 126},
  \href{http://arxiv.org/abs/2012.09873}{{arXiv:2012.09873 [hep-ph]}}.

\bibitem{Butter:2022vkj}
A.~Butter, T.~Heimel, T.~Martini, S.~Peitzsch, and T.~Plehn, {\it {Two
  Invertible Networks for the Matrix Element Method}},
  \href{http://arxiv.org/abs/2210.00019}{{arXiv:2210.00019 [hep-ph]}}.

\bibitem{Heimel:2023mvw}
T.~Heimel, N.~Huetsch, R.~Winterhalder, T.~Plehn, and A.~Butter, {\it
  {Precision-Machine Learning for the Matrix Element Method}},
  \href{http://arxiv.org/abs/2310.07752}{{arXiv:2310.07752 [hep-ph]}}.

\bibitem{Nachman:2020lpy}
B.~Nachman and D.~Shih, {\it {Anomaly Detection with Density Estimation}},
  \href{http://dx.doi.org/10.1103/PhysRevD.101.075042}{Phys. Rev. D {\bfseries
  101} (2020)  075042},
  \href{http://arxiv.org/abs/2001.04990}{{arXiv:2001.04990 [hep-ph]}}.

\bibitem{Hallin:2021wme}
A.~Hallin, J.~Isaacson, G.~Kasieczka, C.~Krause, B.~Nachman, T.~Quadfasel,
  M.~Schlaffer, D.~Shih, and M.~Sommerhalder, {\it {Classifying Anomalies
  THrough Outer Density Estimation (CATHODE)}},
  \href{http://arxiv.org/abs/2109.00546}{{arXiv:2109.00546 [hep-ph]}}.

\bibitem{Raine:2022hht}
J.~A. Raine, S.~Klein, D.~Sengupta, and T.~Golling, {\it {CURTAINs for your
  sliding window: Constructing unobserved regions by transforming adjacent
  intervals}},  \href{http://dx.doi.org/10.3389/fdata.2023.899345}{Front. Big
  Data {\bfseries 6} (2023)  899345},
  \href{http://arxiv.org/abs/2203.09470}{{arXiv:2203.09470 [hep-ph]}}.

\bibitem{Hallin:2022eoq}
A.~Hallin, G.~Kasieczka, T.~Quadfasel, D.~Shih, and M.~Sommerhalder, {\it
  {Resonant anomaly detection without background sculpting}},
  \href{http://dx.doi.org/10.1103/PhysRevD.107.114012}{Phys. Rev. D {\bfseries
  107} (2023) 11, 114012},
  \href{http://arxiv.org/abs/2210.14924}{{arXiv:2210.14924 [hep-ph]}}.

\bibitem{Golling:2022nkl}
T.~Golling, S.~Klein, R.~Mastandrea, and B.~Nachman, {\it {Flow-enhanced
  transportation for anomaly detection}},
  \href{http://dx.doi.org/10.1103/PhysRevD.107.096025}{Phys. Rev. D {\bfseries
  107} (2023) 9, 096025},
  \href{http://arxiv.org/abs/2212.11285}{{arXiv:2212.11285 [hep-ph]}}.

\bibitem{Sengupta:2023xqy}
D.~Sengupta, S.~Klein, J.~A. Raine, and T.~Golling, {\it {CURTAINs Flows For
  Flows: Constructing Unobserved Regions with Maximum Likelihood Estimation}},
   \href{http://arxiv.org/abs/2305.04646}{{arXiv:2305.04646 [hep-ph]}}.

\bibitem{Czakon:2013goa}
M.~Czakon, P.~Fiedler, and A.~Mitov, {\it {Total Top-Quark Pair-Production
  Cross Section at Hadron Colliders Through ${\cal O}(\alpha_s^4)$}},
  \href{http://dx.doi.org/10.1103/PhysRevLett.110.252004}{Phys. Rev. Lett.
  {\bfseries 110} (2013)  252004},
\href{http://arxiv.org/abs/1303.6254}{{arXiv:1303.6254 [hep-ph]}}.
%%CITATION = ARXIV:1303.6254;%%.

\bibitem{Czakon:2015owf}
M.~Czakon, D.~Heymes, and A.~Mitov, {\it {High-precision differential
  predictions for top-quark pairs at the LHC}},
  \href{http://dx.doi.org/10.1103/PhysRevLett.116.082003}{Phys. Rev. Lett.
  {\bfseries 116} (2016) 8, 082003},
\href{http://arxiv.org/abs/1511.00549}{{arXiv:1511.00549 [hep-ph]}}.
%%CITATION = ARXIV:1511.00549;%%.

\bibitem{Catani:2019iny}
S.~Catani, S.~Devoto, M.~Grazzini, S.~Kallweit, J.~Mazzitelli, and H.~Sargsyan,
  {\it {Top-quark pair hadroproduction at next-to-next-to-leading order in
  QCD}},  \href{http://dx.doi.org/10.1103/PhysRevD.99.051501}{Phys. Rev.
  {\bfseries D99} (2019) 5, 051501},
\href{http://arxiv.org/abs/1901.04005}{{arXiv:1901.04005 [hep-ph]}}.
%%CITATION = ARXIV:1901.04005;%%.

\bibitem{Catani:2019hip}
S.~Catani, S.~Devoto, M.~Grazzini, S.~Kallweit, and J.~Mazzitelli, {\it
  {Top-quark pair production at the LHC: Fully differential QCD predictions at
  NNLO}},  \href{http://dx.doi.org/10.1007/JHEP07(2019)100}{JHEP {\bfseries 07}
  (2019)  100},
\href{http://arxiv.org/abs/1906.06535}{{arXiv:1906.06535 [hep-ph]}}.
%%CITATION = ARXIV:1906.06535;%%.

\bibitem{Bernreuther:2008md}
W.~Bernreuther, M.~Fucker, and Z.-G. Si, {\it {Weak interaction corrections to
  hadronic top quark pair production: Contributions from quark-gluon and b
  anti-b induced reactions}},
  \href{http://dx.doi.org/10.1103/PhysRevD.78.017503}{Phys. Rev. {\bfseries
  D78} (2008)  017503},
\href{http://arxiv.org/abs/0804.1237}{{arXiv:0804.1237 [hep-ph]}}.
%%CITATION = ARXIV:0804.1237;%%.

\bibitem{Kuhn:2006vh}
J.~H. K{\"u}hn, A.~Scharf, and P.~Uwer, {\it {Electroweak effects in top-quark
  pair production at hadron colliders}},
  \href{http://dx.doi.org/10.1140/epjc/s10052-007-0275-x}{Eur. Phys. J.
  {\bfseries C51} (2007)  37},
\href{http://arxiv.org/abs/hep-ph/0610335}{{arXiv:hep-ph/0610335 [hep-ph]}}.
%%CITATION = HEP-PH/0610335;%%.

\bibitem{Hollik:2011ps}
W.~Hollik and D.~Pagani, {\it {The electroweak contribution to the top quark
  forward-backward asymmetry at the Tevatron}},
  \href{http://dx.doi.org/10.1103/PhysRevD.84.093003}{Phys. Rev. {\bfseries
  D84} (2011)  093003},
\href{http://arxiv.org/abs/1107.2606}{{arXiv:1107.2606 [hep-ph]}}.
%%CITATION = ARXIV:1107.2606;%%.

\bibitem{Gutschow:2018tuk}
C.~G\"utschow, J.~M. Lindert, and M.~Sch\"onherr, {\it {Multi-jet merged
  top-pair production including electroweak corrections}},
  \href{http://dx.doi.org/10.1140/epjc/s10052-018-5804-2}{Eur. Phys. J. C
  {\bfseries 78} (2018) 4, 317},
  \href{http://arxiv.org/abs/1803.00950}{{arXiv:1803.00950 [hep-ph]}}.

\bibitem{Frederix:2021zsh}
R.~Frederix, I.~Tsinikos, and T.~Vitos, {\it {Probing the spin correlations of
  $t{\bar{t}} $ production at NLO QCD+EW}},
  \href{http://dx.doi.org/10.1140/epjc/s10052-021-09612-9}{Eur. Phys. J. C
  {\bfseries 81} (2021) 9, 817},
  \href{http://arxiv.org/abs/2105.11478}{{arXiv:2105.11478 [hep-ph]}}.

\bibitem{Czakon:2017wor}
M.~Czakon, D.~Heymes, A.~Mitov, D.~Pagani, I.~Tsinikos, and M.~Zaro, {\it
  {Top-pair production at the LHC through NNLO QCD and NLO EW}},
  \href{http://dx.doi.org/10.1007/JHEP10(2017)186}{JHEP {\bfseries 10} (2017)
  186},
\href{http://arxiv.org/abs/1705.04105}{{arXiv:1705.04105 [hep-ph]}}.
%%CITATION = ARXIV:1705.04105;%%.

\bibitem{Gao:2012ja}
J.~Gao, C.~S. Li, and H.~X. Zhu, {\it {Top Quark Decay at Next-to-Next-to
  Leading Order in QCD}},
  \href{http://dx.doi.org/10.1103/PhysRevLett.110.042001}{Phys. Rev. Lett.
  {\bfseries 110} (2013) 4, 042001},
\href{http://arxiv.org/abs/1210.2808}{{arXiv:1210.2808 [hep-ph]}}.
%%CITATION = ARXIV:1210.2808;%%.

\bibitem{Brucherseifer:2013iv}
M.~Brucherseifer, F.~Caola, and K.~Melnikov, {\it {$\mathcal O(\alpha_s^2)$
  corrections to fully-differential top quark decays}},
  \href{http://dx.doi.org/10.1007/JHEP04(2013)059}{JHEP {\bfseries 04} (2013)
  059},
\href{http://arxiv.org/abs/1301.7133}{{arXiv:1301.7133 [hep-ph]}}.
%%CITATION = ARXIV:1301.7133;%%.

\bibitem{Gao:2017goi}
J.~Gao and A.~S. Papanastasiou, {\it {Top-quark pair-production and decay at
  high precision}},  \href{http://dx.doi.org/10.1103/PhysRevD.96.051501}{Phys.
  Rev. {\bfseries D96} (2017) 5, 051501},
\href{http://arxiv.org/abs/1705.08903}{{arXiv:1705.08903 [hep-ph]}}.
%%CITATION = ARXIV:1705.08903;%%.

\bibitem{Behring:2019iiv}
A.~Behring, M.~Czakon, A.~Mitov, A.~S. Papanastasiou, and R.~Poncelet, {\it
  {Higher order corrections to spin correlations in top quark pair production
  at the LHC}},  \href{http://dx.doi.org/10.1103/PhysRevLett.123.082001}{Phys.
  Rev. Lett. {\bfseries 123} (2019) 8, 082001},
\href{http://arxiv.org/abs/1901.05407}{{arXiv:1901.05407 [hep-ph]}}.
%%CITATION = ARXIV:1901.05407;%%.

\bibitem{Czakon:2020qbd}
M.~Czakon, A.~Mitov, and R.~Poncelet, {\it {NNLO QCD corrections to leptonic
  observables in top-quark pair production and decay}},
  \href{http://dx.doi.org/10.1007/JHEP05(2021)212}{JHEP {\bfseries 05} (2021)
  212},  \href{http://arxiv.org/abs/2008.11133}{{arXiv:2008.11133 [hep-ph]}}.

\bibitem{Bevilacqua:2010qb}
G.~Bevilacqua, M.~Czakon, A.~van Hameren, C.~G. Papadopoulos, and M.~Worek,
  {\it {Complete off-shell effects in top quark pair hadroproduction with
  leptonic decay at next-to-leading order}},
  \href{http://dx.doi.org/10.1007/JHEP02(2011)083}{JHEP {\bfseries 02} (2011)
  083},
\href{http://arxiv.org/abs/1012.4230}{{arXiv:1012.4230 [hep-ph]}}.
%%CITATION = ARXIV:1012.4230;%%.

\bibitem{Denner:2010jp}
A.~Denner, S.~Dittmaier, S.~Kallweit, and S.~Pozzorini, {\it {NLO QCD
  corrections to WWbb production at hadron colliders}},
  \href{http://dx.doi.org/10.1103/PhysRevLett.106.052001}{Phys. Rev. Lett.
  {\bfseries 106} (2011)  052001},
\href{http://arxiv.org/abs/1012.3975}{{arXiv:1012.3975 [hep-ph]}}.
%%CITATION = ARXIV:1012.3975;%%.

\bibitem{Denner:2012yc}
A.~Denner, S.~Dittmaier, S.~Kallweit, and S.~Pozzorini, {\it {NLO QCD
  corrections to off-shell top-antitop production with leptonic decays at
  hadron colliders}},  \href{http://dx.doi.org/10.1007/JHEP10(2012)110}{JHEP
  {\bfseries 10} (2012)  110},
\href{http://arxiv.org/abs/1207.5018}{{arXiv:1207.5018 [hep-ph]}}.
%%CITATION = ARXIV:1207.5018;%%.

\bibitem{Heinrich:2013qaa}
G.~Heinrich, A.~Maier, R.~Nisius, J.~Schlenk, and J.~Winter, {\it {NLO QCD
  corrections to $W^{+} W^{-}b\bar{b}$ production with leptonic decays in the
  light of top quark mass and asymmetry measurements}},
  \href{http://dx.doi.org/10.1007/JHEP06(2014)158}{JHEP {\bfseries 06} (2014)
  158},
\href{http://arxiv.org/abs/1312.6659}{{arXiv:1312.6659 [hep-ph]}}.
%%CITATION = ARXIV:1312.6659;%%.

\bibitem{Frederix:2013gra}
R.~Frederix, {\it {Top Quark Induced Backgrounds to Higgs Production in the
  $WW^{(*)}\to ll\nu\nu$ Decay Channel at Next-to-Leading-Order in QCD}},
  \href{http://dx.doi.org/10.1103/PhysRevLett.112.082002}{Phys. Rev. Lett.
  {\bfseries 112} (2014) 8, 082002},
\href{http://arxiv.org/abs/1311.4893}{{arXiv:1311.4893 [hep-ph]}}.
%%CITATION = ARXIV:1311.4893;%%.

\bibitem{Cascioli:2013wga}
F.~Cascioli, S.~Kallweit, P.~Maierh{\"o}fer, and S.~Pozzorini, {\it {A unified
  NLO description of top-pair and associated Wt production}},
  \href{http://dx.doi.org/10.1140/epjc/s10052-014-2783-9}{Eur. Phys. J.
  {\bfseries C74} (2014) 3, 2783},
\href{http://arxiv.org/abs/1312.0546}{{arXiv:1312.0546 [hep-ph]}}.
%%CITATION = ARXIV:1312.0546;%%.

\bibitem{Bevilacqua:2015qha}
G.~Bevilacqua, H.~B. Hartanto, M.~Kraus, and M.~Worek, {\it {Top Quark Pair
  Production in Association with a Jet with Next-to-Leading-Order QCD Off-Shell
  Effects at the Large Hadron Collider}},
  \href{http://dx.doi.org/10.1103/PhysRevLett.116.052003}{Phys. Rev. Lett.
  {\bfseries 116} (2016) 5, 052003},
\href{http://arxiv.org/abs/1509.09242}{{arXiv:1509.09242 [hep-ph]}}.
%%CITATION = ARXIV:1509.09242;%%.

\bibitem{NNPDF:2017mvq}
NNPDF, R.~D. Ball {\em et al.}, {\it {Parton distributions from high-precision
  collider data}},
  \href{http://dx.doi.org/10.1140/epjc/s10052-017-5199-5}{Eur. Phys. J. C
  {\bfseries 77} (2017) 10, 663},
  \href{http://arxiv.org/abs/1706.00428}{{arXiv:1706.00428 [hep-ph]}}.

\bibitem{Frixione:2007nw}
S.~Frixione, P.~Nason, and G.~Ridolfi, {\it {A Positive-weight
  next-to-leading-order Monte Carlo for heavy flavour hadroproduction}},
  \href{http://dx.doi.org/10.1088/1126-6708/2007/09/126}{JHEP {\bfseries 09}
  (2007)  126},  \href{http://arxiv.org/abs/0707.3088}{{arXiv:0707.3088
  [hep-ph]}}.

\bibitem{Jezo:2016ujg}
T.~Je\v{z}o, J.~M. Lindert, P.~Nason, C.~Oleari, and S.~Pozzorini, {\it {An
  NLO+PS generator for $t\bar{t}$ and $Wt$ production and decay including
  non-resonant and interference effects}},
  \href{http://dx.doi.org/10.1140/epjc/s10052-016-4538-2}{Eur. Phys. J. C
  {\bfseries 76} (2016) 12, 691},
  \href{http://arxiv.org/abs/1607.04538}{{arXiv:1607.04538 [hep-ph]}}.

\bibitem{Jezo:2023rht}
T.~Je\v{z}o, J.~M. Lindert, and S.~Pozzorini, {\it {Resonance-aware NLOPS
  matching for off-shell $ t\overline{t} $ + tW production with semileptonic
  decays}},  \href{http://dx.doi.org/10.1007/JHEP10(2023)008}{JHEP {\bfseries
  10} (2023)  008},  \href{http://arxiv.org/abs/2307.15653}{{arXiv:2307.15653
  [hep-ph]}}.

\bibitem{Frixione:2007zp}
S.~Frixione, E.~Laenen, P.~Motylinski, and B.~R. Webber, {\it {Angular
  correlations of lepton pairs from vector boson and top quark decays in Monte
  Carlo simulations}},
  \href{http://dx.doi.org/10.1088/1126-6708/2007/04/081}{JHEP {\bfseries 04}
  (2007)  081},
  \href{http://arxiv.org/abs/hep-ph/0702198}{{arXiv:hep-ph/0702198}}.

\bibitem{Nason:2004rx}
P.~Nason, {\it {A New method for combining NLO QCD with shower Monte Carlo
  algorithms}},  \href{http://dx.doi.org/10.1088/1126-6708/2004/11/040}{JHEP
  {\bfseries 11} (2004)  040},
  \href{http://arxiv.org/abs/hep-ph/0409146}{{arXiv:hep-ph/0409146}}.

\bibitem{Frixione:2007vw}
S.~Frixione, P.~Nason, and C.~Oleari, {\it {Matching NLO QCD computations with
  Parton Shower simulations: the POWHEG method}},
  \href{http://dx.doi.org/10.1088/1126-6708/2007/11/070}{JHEP {\bfseries 11}
  (2007)  070},  \href{http://arxiv.org/abs/0709.2092}{{arXiv:0709.2092
  [hep-ph]}}.

\bibitem{Alioli:2010xd}
S.~Alioli, P.~Nason, C.~Oleari, and E.~Re, {\it {A general framework for
  implementing NLO calculations in shower Monte Carlo programs: the POWHEG
  BOX}},  \href{http://dx.doi.org/10.1007/JHEP06(2010)043}{JHEP {\bfseries 06}
  (2010)  043},  \href{http://arxiv.org/abs/1002.2581}{{arXiv:1002.2581
  [hep-ph]}}.

\bibitem{Jezo:2015aia}
T.~Je\v{z}o and P.~Nason, {\it {On the Treatment of Resonances in
  Next-to-Leading Order Calculations Matched to a Parton Shower}},
  \href{http://dx.doi.org/10.1007/JHEP12(2015)065}{JHEP {\bfseries 12} (2015)
  065},  \href{http://arxiv.org/abs/1509.09071}{{arXiv:1509.09071 [hep-ph]}}.

\bibitem{FM2022}
Y.~Lipman, R.~T.~Q. Chen, H.~Ben-Hamu, M.~Nickel, and M.~Le, {\it Flow matching
  for generative modeling},
  \href{http://arxiv.org/abs/2210.02747}{{arXiv:2210.02747 [cs.LG]}}.

\bibitem{Bellagente:2021yyh}
M.~Bellagente, M.~Hau\ss{}mann, M.~Luchmann, and T.~Plehn, {\it {Understanding
  Event-Generation Networks via Uncertainties}},
  \href{http://dx.doi.org/10.21468/SciPostPhys.13.1.003}{SciPost Phys.
  {\bfseries 13} (2022)  003},
  \href{http://arxiv.org/abs/2104.04543}{{arXiv:2104.04543 [hep-ph]}}.

\bibitem{OTCFM}
A.~Tong, N.~Malkin, G.~Huguet, Y.~Zhang, J.~Rector-Brooks, K.~Fatras, G.~Wolf,
  and Y.~Bengio, {\it Improving and generalizing flow-based generative models
  with minibatch optimal transport},
  \href{http://arxiv.org/abs/2302.00482}{{arXiv:2302.00482 [cs.LG]}}.

\end{thebibliography}\endgroup
\end{document}